\documentclass[]{interact}

\usepackage{epstopdf}
\usepackage[caption=false]{subfig}
\usepackage{hyperref}
\usepackage{xcolor}
\usepackage{lmodern}

\usepackage[natbibapa,nodoi]{apacite}
\theoremstyle{plain}

\theoremstyle{definition}

\theoremstyle{remark}

\begin{document}

\title{Towards a feminist understanding of digital platform work}

\author{
\name{Clara Punzi\textsuperscript{a}\thanks{Corresponding author. Email: clara.punzi@sns.it}}
\affil{\textsuperscript{a}Scuola Normale Superiore, Pisa, Italy}
}

\maketitle

\begin{abstract}
The rapid growth of the digital platform economy is transforming labor markets, offering new employment opportunities with promises of flexibility and accessibility. However, these benefits often come at the expense of increased economic exploitation, occupational segregation, and deteriorating working conditions. Research highlights that algorithmic management disproportionately impacts marginalized groups, reinforcing gendered and racial inequalities while deepening power imbalances within capitalist systems. This study seeks to elucidate the complex nature of digital platform work by drawing on feminist theories that have historically scrutinized and contested the structures of power within society, especially in the workplace. It presents a framework focused on four key dimensions to lay a foundation for future research: (i) precarity and exploitation, (ii) surveillance and control, (iii) blurring employment boundaries, and (iv) colonial legacies. It advocates for participatory research, transparency in platform governance, and structural changes to promote more equitable conditions for digital platform workers.

\end{abstract}

\section*{Introduction}

The use of digital technology to provide services is becoming ubiquitous and is quickly transforming working and everyday lives, resulting in new relationships between digital technology, society, and the physical space \citep{Elwood2020,Webster2022}. The platform economy provides novel opportunities for employment and alternative methods of consumption, which have specific effects on the access to work, occupational segregation and precarity, the organization of work and other working conditions \citep{Cirillo2023,Hoang2020}. In particular, emerging technologies often have different impacts across occupations and skills, frequently exacerbating inequalities \citep{Pianta2020}. Indeed, recent economic analyses stress that digital platforms may be characterized as capitalist monopolies that accumulate power by exerting control over workers, governments, competitors, and clients \citep{Coveri2022}.

Workers in the digital platform economy, such as annotators and digital content moderators, occupy the lowest position in the professional ladder that defines occupations in the field of Artificial Intelligence (AI), with their activities remaining mostly invisible and unaccredited \cite{Muldoon2023Matrix}.
\citet{Klein2024} noted that the professional hierarchy aligns with gendered, racial, and colonial hierarchies: individuals in the Global North hold prestigious and well-paid positions, whereas individuals in the Global South occupy lower-ranking ones. 
Analogously, the workforce in the Global North also reproduces a professional hierarchy based on marginalized social categories and dynamics, including gender, age, and migratory status \cite{vanDoorn2024,altenried2022platforms,Rodriguez2022}.
This suggest that the digital economy is not ahistorical nor a neutral outcome of technological advances \citep{Webster2020}. 
The general trend observed in the digital platform economy reveals a dual interaction with marginalization. First, studies indicate that marginalized groups are disproportionately engaged in digital labor \cite{Munoz2022}, suggesting that platform work can serve simultaneously as a point of entry, emergence, and segregation for individuals already facing discrimination. Second, the algorithmic management systems governing digital platforms have the potential to reinforce and amplify existing biases tied to social categories and power asymmetries in the traditional labor market. This dynamic not only reproduces structural inequalities but also creates new, and potentially more radical, forms of marginalization within the digital space \cite{Pianta2020,smith2021good,Tubaro2022}.

Building on this evidence, this study seeks to provide a deeper understanding of the complexities of digital labor by drawing on feminist theories that have long examined and challenged the ways in which power is structured and sustained in society, particularly in the workplace. We argue that these perspectives may help reveal the socio-economic inequalities that shape the lived experiences of women and other marginalized groups in the platform economy. As a result, we propose a framework based on four key feminist dimensions that entail the contributions of feminist theory to this debate and offer a foundation for future research: (i) precarity, informality, and economic exploitation; (ii) surveillance and control; (iii) the blurring of employment boundaries; and (iv) the persistence of colonial structures.


The study is structured into two interconnected parts. In \autoref{sec:techWork}, we briefly outline the political and philosophical discussions on the relationship between technology and work, emphasizing the ambivalent nature of automation in reshaping labor, as well as the resulting power dynamics. While first addressing technology broadly, we then concentrate on the innovations associated with AI in the digital age to present the new landscape of digital platform work.
In the second part (\autoref{sec:feministDim}), we analyze relevant feminist theories that offer essential tools for a thorough examination of the digital platform economy. Subsequently, we outline our primary contributions, which consist of the identification of four key dimensions of analysis that we assert could encompass the contributions of feminist theories to this discourse and offer direction for future inquiry. 
Finally, \autoref{sec:conclusion} presents a synthesis of findings and outlines potential avenues for future investigation.

 
\section{Technology and work: power dynamics beyond emerging inequalities}
\label{sec:techWork}

The interplay between technology and work has long been central to political and philosophical debates on economic and social structures. From the Industrial Revolution, where mechanization and factory systems displaced artisanal work and centralized production, to the emergence of automation, technological advancements have consistently altered the balance between capital and labor \cite{Pianta2020}. Within a wider process of ``creative destruction'' \cite{Schumpeter1976}, such changes have created new employment opportunities while rendering others redundant or requiring readaptation, thereby raising questions regarding the value of human labor and promoting discussions about the potential, at times necessary, restructuring of the political and economic structure of society. 

The ambivalent nature of technology and its impact on labor, either empowering or exploitative \cite{Richardson2016}, has been addressed by many political economists and philosophers.  
\citet{Marx1961} viewed technology as a tool of capitalist control which, while increasing productivity, also alienates workers and consolidates power in the hands of capitalists by de-skilling labor and intensifying exploitation. By keeping the means of production, including automated technologies, under their ownership, capitalists concentrate the profits generated by increased productivity, further exacerbating economic inequality and class divisions.
He claimed that unemployment rises when technological advancements displace labor faster than capital expansion creates new jobs \cite{Marx1961}, suggesting that technological unemployment is inherently linked to the character of capitalist production \cite{Pianta2020}.
At the same time, he established the foundation for subsequent critiques of capitalism that envisage a future in which, under a different socio-economic system, technological advancement would empower individuals to overcome alienated labor, converting work from a necessity into a creative, self-actualizing activity \cite{Marx1973,Keynes2010}. 
This duality was further explored, among others, by the Frankfurt School philosopher Herbert Marcuse. In his works, he claimed that, on one hand, technological automation of labor could be the condition for a process of emancipating social life from alienation \cite{Marcuse1955}. 
Nevertheless, he subsequently contends that, inside capitalist frameworks, technology often transforms into an instrument of domination and subjugation of knowledge, supplanting terror in organizing social control via its unprecedented efficiency and higher standards of living \cite{Marcuse1964}.

Despite Marx's commitment to the liberation of work from exploitation and the restoration of its dignity in an unalienated form, influenced by Hegel's exaltation of work as a prototypically human  endeavor \cite{Hegel1991}, alternative traditions pursued a divergent stand towards envisioning a world liberated from labor, possibly enabled by technological progress \cite{Rifkin1995,Aronowitz1998,Weeks2011}. One noteworthy example is the Autonomist Marxists political theory, which call not for a liberation \textit{of} work but for a liberation \textit{from} work \cite{Virno1996}. Importantly, this refusal (and potential abolition) of work should not be understood as the blind negation of activity, but rather as a valorization of human activities that have escaped from labor's domination \cite{Berardi2009,Negri1979}. Within the Autonomist Marxist tradition, technology plays a crucial role. According to \citet{Bifo2018}, progress in automation has the potential to completely liberate humans from labor. However, he acknowledges that this venture is being continuously hindered by the capitalist societal structure, alongside cultural resistance, imbalances in economic development in different areas of the planet, the effects of global competition on the profit economy, and, ultimately, the contradictory role played by the industrial workers' movement, including communist parties and labor unions. As a result, technological progress has led not to liberation from labor, but to increased precarity 
and a widespread sense of subjective and political impotence which, coupled with social depletion, stem from the uncontrollable intensification of information stimulation and simulation.
An analogous liberatory potential in technology was articulated by the anarchist theorist Murray Bookchin, who claimed that the technological revolution had established the objective preconditions for a society devoid of class domination, exploitation, labor, or material deprivation. At the same time, he recognized the adverse, socially regressive aspects of technological advancement and contended that only a decentralised society, such as one founded on communes and affinity groups, will be able to effectively harness technology for this goal \cite{Bookchin1986}.


The historical trajectory of technology underlines its tendency to perpetuate inequities and power asymmetries, underscoring the need for critical examination of its deployment and control within specific labor settings.
Indeed, work is always situated inside a distinct historical and economic framework that delineates power dynamics and societal roles, along with the valuation of workers' endeavors and products \cite{Cukier2018}.
Moreover, digital spaces, including digital labor platforms, are shaped by social relations \citep{Pianta2020}, meaning they reflect and perpetuate the complex interplay of existing and evolving forces dynamics \citep{Elwood2020,Webster2022}, such as gender-based power relationships and colonial frameworks of modernity and progress \citep{Rani2022}.
Therefore, to accurately understand and evaluate the real effects of technology on the quantity and quality of work, it is essential to consider the specific temporal and spatial contexts, alongside the influence of growth models, social relationships, institutional structures, and the overarching policy frameworks of various countries \citep{Pianta2020}. 

\subsection{Trajectories of work in the digital era}
The shift into the digital era, and the recent the disruption of Generative AI, has been accompanied by mounting scrutiny over the changes it will bring on the workforce and on the nature of work itself, with perspectives ranging from fears of job loss or exploitative digital labor to the potential for greater flexibility and the emergence of innovative, creative employment opportunities \cite{Fayard2021}. 
For instance, it has raised expectations for novel avenues of empowerment, with international development agencies \citep{UNWomen2020} that have strongly advocated for leveraging digital technologies as a strategic method to attain sustainable, inclusive, and equitable economic growth \citep{Rani2022}. A diverse array of jobs has also been boosted, including software developers, programmers, market sellers and micro-entrepreneurs who utilize digital tools to connect with clients \cite{Rani2022}. Embraced in this category are also digital platform workers, which encompass both web-based crowdworkers (e.g., persons executing tasks via platforms such as Amazon Mechanical Turk) and offline platform workers or gig-workers (e.g., those engaged in food delivery, ride-hailing, domestic and care work managed through digital platforms) \cite{Stefano2015}. Nonetheless, the quality of these jobs and their effects on employees have yet to be thoroughly evaluated. Yet, overall, the broad category of digital platform workers is emerging as a non-standard form of low-paying employment that fall outside the scope of labor contracts and is not usually covered by unions \cite{Pianta2020}.

As increasing studies and empirical evidences demonstrate, while digital platform work may provide some individuals with eased access to the job market, supplementary earnings possibilities, and greater flexibility in managing their working hours, it may also lead to precarity, technology-enabled surveillance, aggravated power disparities, and diminished transparency in decision-making, particularly if left unregulated.
The impact of technology varies across occupations and skills, resulting in a polarized employment structure that favors those at higher positions while negatively affecting a growing proportion of unskilled workers \cite{Pianta2020}. This dynamics further exacerbated the role of hierarchies in the workplace in terms of power, control over work, and remuneration \cite{Pianta2020}.
In other words, we can say that the emergence of new forms of employment mediated by digital platforms signifies a novel iteration of the ambivalent interplay between technology and work, as recent legal documents are also starting to claim \citep{EUDirectiveWork2024}. 
While technology, in a broader sense, reconfigures labor power dynamics by amplifying employers’ authority through greater control over production processes and reducing dependence on skilled labor, in contemporary workplaces, digital technologies such as AI, surveillance tools, and algorithmic management reinforce this tendency. Employers now have unprecedented access to data about workers’ productivity and behavior, facilitating micromanagement and diminishing workplace privacy. Algorithms increasingly mediate tasks like scheduling, monitoring, and performance evaluation, often lacking transparency and human involvement. 
Concealed under the facade of automated decision-making and neutral service provision, these systems reveal a discernible exertion of social power \cite{Muldoon23}.
This asymmetry of knowledge and control strengthens employer power while limiting workers' capacity to bargain, organize collectively, or resist unfair practices.

In her seminal work, \citet{Zuboff2019} coins the term \textit{surveillance capitalism} to describe the current process of technological transformation where dominant actors in capitalism have established monopolistic control over information-based activities. She points out that they maintain their authority by extracting and claiming ownership of all accessible data from individuals, forecasting and manipulating their behavior as consumers, workers, and citizens, selling this information to businesses that customize their offerings based on specific consumer profiles, and establishing global platforms to create innovative models for flexible work arrangements. Zuboff also claims that surveillance capitalism seek to limit human nature and restrict the freedom and capacity to act of individuals, posing a significant challenge to democracy. Indeed, \citet{Pianta2020} noted that, while the advent of new information and communications technologies has significantly increased the potential for both enhancing and deskilling work, this potential has often been realized through the implementation of these technologies and the organization of work in a way that prioritizes reducing labor and gain greater control over workers.

The transformative potential of digital technology can be observed in the substantial monopolistic power lately attained by the major players of AI, in particular digital platforms, as well as its repercussion on the evolving structure of the workforce.
According to \citet{Huws2014}, this monopolistic power is the result of digital innovation coupled with a decline in the regulatory capacity of governments, which have both facilitated the concentration of capital in specific industries and geographies.
Indeed, \citet{Coveri2022} suggested that digital technologies derive their power from a sort of \textit{digital} (or \textit{virtual}) \textit{colonization}. Similarly, \citet{Couldry2019} argue that data colonization is an extension of the capitalist mode of operation that not only turns human activity into labor for profit, but also converts life itself into data for economic exploitation. Digital colonialism, in contrast to traditional colonialism, exhibits a more nuanced and widespread presence. Although it may not be evident in the form of visible violence, its effects, such as surveillance, commodification, and marginalization, still have a profound detrimental impact \citep{Couldry2019}.
In line with the same historical parallel, the colonial legacy has been suggested as the primary cause for the uneven economic development on a global level \cite{Kyrylych2013}, as further laid out by the \textit{law of uneven development} \citep{Trotsky1977}. This principle suggests that there is a tendency for a hierarchical division of labor between different regions, corresponding to the vertical division of labor within a company. This means that a few key geographical areas have high-level decision-making roles, while the rest of the world is confined to lower levels of economic activity and income \citep{Hymer1982}. The advantages of the colonizing players can be attributed to a specific set of criteria, primarily geographical position, resource potential, labor force potential, and availability of capital \citep{Kyrylych2013}.

\section{A feminist analysis of the digital platform economy}
\label{sec:feministDim}
The analysis in \autoref{sec:techWork} suggests that the integration of technological advancements, especially AI, into the workplace is driving a new phase of capitalism. This phase is marked by aggravated power asymmetries at various levels (such as geographical, economic, racial, gender, and knowledge-related), as well as significant levels of unemployment and socio-economic inequality.
Although the current evolution has given rise to the unique characteristics we have previously examined, it is important to note that these issues are not historically novel. Therefore, it could prove valuable to reclaim previous critical studies in order to situate this new phenomenon in its historical context and then adjust them in accordance with the current paradigm.
In particular, feminism, as a broad and diverse intellectual and political movement, has always been deeply concerned with power asymmetries and inequalities. Indeed, different feminist theories have critically examined and challenged the ways in which various forms of power are structured and perpetuated in society, leading to socio-economic inequalities that shape the lived experiences of women and other marginalized groups.

\textit{Feminist economics} (FE) is a noteworthy school of thought and political action that challenges the core principles of neoclassical economics. It emerged prominently in the 1990s, although its roots may be traced back to the mid 19th century. It encompasses a diverse body of perspectives \cite{Carrasco2014} that converge in a shared perception of the economy as a means to bring about social change and establish an economy that operates on the basis of justice and equality \cite{Agenjo2019}. Overall, the aim of FE is to transcend a reductionist, biased, and hierarchical perspective and to develop novel economic principles that prioritize the everyday experiences of individuals.
Furthermore, \citet{Agenjo2019} identify a set of key principles shared by all FE approaches, namely: (i) a deeper understanding of the mechanisms that sustain life and social support, as well as the mechanisms of inclusion and exclusion, and recognition of gender as a fundamental category; (ii) recognition of the significance of unpaid domestic work and caregiving; (iii) utilization of human well-being as a metric for economic achievement; (iv) incorporation of intersectional analysis, considering the various social dimensions that shape individuals' lives and identities; (v) acknowledgment of the value of social activism and the necessity of incorporating ethical evaluations in economic analysis.

As already stressed, feminist thoughts offer valuable insights into the complex and multifaceted aspects of work with and through digital technologies.
They provide frameworks for examining the ambivalent nature of digital labor, encompassing both its opportunity for empowering workers and liberating them from domination, as well as its potential for aggravated exploitation and alienation \citep{Richardson2016}.
In the following, we examine relevant feminist theories that provide critical instruments for analyzing the production, perpetuation and reinforcement of social categories, hierarchies and power dynamics within the specific context of the digital platform economy. As a result, we delineate four key categories of analysis that we claim might encapsulate the contributions of feminist theories to this debate and provide guidance for future exploration. 

\subsection{Precarity, informality and economic exploitation}


Historically, several feminist theories have addressed issues of informal labor, precarity, and economic exploitation, focusing on how these forms of labor are gendered and tied to broader systems of inequality. 
For instance, \textit{Social Reproduction Theory} (SRT) examines how capitalist systems rely not only on waged labor but also on the unpaid and undervalued labor involved in maintaining and reproducing the labor force. This labor includes a wide range of activities, such as caregiving, domestic work, education, and emotional support, that are essential to sustaining workers and their ability to participate in the economy, yet are often relegated to women and marginalized groups. According to \citet{Bhattacharya2017}, by prioritizing profit over people, capitalism intensifies exploitation both in the formal labor market and in the realm of social reproduction, where the essential work of caring for society is often carried out under precarious conditions.
\citet{Federici1984} employ SRT to explain the Marxist notion of primitive accumulation. They demonstrate that capitalism was not solely established through the enclosure of land, but also through the dispossession resulting from imperial and colonial endeavors as well as through the expropriation and exclusion of certain groups of people, such as housewives, from the arenas where value is generated or appropriated.
Similarly, \citet{Mies1982}' examination of home-based work challenges those theories that advocate for a clear distinction between production and reproduction. It highlights how the domestication of women's labor has consistently obscured the sources of value, both by concealing women's productive contributions to the market and by undervaluing those contributions as non-value-producing.

More recently, feminist scholars such as Ursula Huws highlight how digital labor markets exploit workers by treating them as independent contractors, thus avoiding the responsibilities of traditional employers. Huws is closely associated with \textit{socialist feminism}, which encompasses a variety of theories that build upon Marxist feminism's argument for the role of capitalism in the oppression of women and radical feminism's theory of the role of gender and the patriarchy. The objective of socialist feminism is to elucidate the complex interaction between capitalism and patriarchy, and how these systems jointly contribute to the oppression of women, particularly in the context of labor and economic exploitation \citep{Tong2018}.
Regarded as a prominent feminist political economist in Europe and beyond, Huws has been studying the issue of social isolation associated with telehomework since the 1980s \citep{IntoTheBlackBox}. She has also focused on the atomization of society and the potential negative impact of new technologies on work-life balance and family dynamics, with a specific emphasis on women's work \citep{Huws2003,Huws2014}.  At the start of the new millennium, her interest shifted towards digital labor. Indeed, she is renowned for coining the word ``cybertariat'', which combines ``cyber'' and ``proletariat'', to describe a working class engaged in repetitive, unskilled, and low-paid digital labor, such as moderating online social media platforms.

\citet{Surie2023}'s edited volume brings together accounts about the effects of platformization in the global South, since they argue that the precarious employment and exploitative working conditions resulting from the digital revolution in global Northern countries share similarities with the long-standing precarization of labor markets in the global South.
They highlight that, although the digital revolution has been identified as the cause of increased insecurity in global Northern countries, most people in the so-called ``Rest" were already working in informal and poor conditions. By citing \citet{Breman2013}, they support the idea that the ``West'' appears to be following the ``Rest'' in terms of the increasing insecurity of work conditions, which contradicts the prevailing belief of the dominant development paradigm. The works collected by \citet{Surie2023} emphasize the continuities with past forms of informal employment together with the disruptions resulting from the incursion of global platforms.
In addition, \citet{Surie2023} argue that the current labor trends in the context of platformization should be more accurately described as new forms of ``informal economy'' rather than a process of ``precarization''. The term ``informal economy'' refers to economic activities performed by workers and economic units that are not adequately covered by formal arrangements, either legally or in practice. It mostly flourishes in a setting characterized by elevated levels of unemployment, insufficient employment, poverty, gender inequality, and precarious work conditions.

In addition, \citet{Huws2019}'s research focuses on examining the gendered dimensions of informality, specifically in the context of domestic labor. This type of platform work, primarily undertaken by women, has received less scholarly attention compared to the more visible forms of work conducted by men in public spaces. Specifically, she claims that technologies, being under the hands of corporations and fueled by profit, are unable to fully deliver on their promise of eliminating need for tedious and hard tasks such as household chores. Moreover, the process of capitalist restructuring generates low-wage employment opportunities for people, simultaneously producing cheap goods. However, the income of workers never quite matches the cost of these goods, leading to a continuous cycle of exploitation. Remarkably, she discovers that the relationship between platform service providers and users does not always involve distinct socioeconomic classes, indicating that novel distributions of power may also cross conventional boundaries.

Tiziana Terranova is not primarily known as a feminist scholar, but rather as a theorist of digital labor. Nevertheless, her work partially intersects with feminist concerns, particularly regarding the commodification of labor and the ways in which certain types of labor, often performed by women, are exploited and undervalued.
Terranova is particularly known for her concept of ``free labor'', which explores how digital platforms and the internet extract value from unpaid or underpaid labor performed by users. Notably, she highlights that this labor is not confined to the traditional workplace but is distributed across society, echoing the Italian \textit{Autonomists}' idea that labor and production have become socialized and diffused throughout all aspects of life \citep{Hardt2005,Tronti2019} and \textit{Autonomist-feminists}' intuition that capitalism exploits not only waged labor in traditional workplaces but also unwaged labor in the home and broader society \citep{DallaCosta2017,Federici1984,Fortunati1995}. In this sense, the digital economy can be seen as a continuation and intensification of the social factory, where the extraction of value from social relations and interactions becomes a central feature of capitalist accumulation \citep{Terranova2012}.

On the same line of thought, \citet{Jarrett2015} introduced the term ``Digital Housewife" to describe <<the actor that emerges from the structures and practices of the ostensibly voluntary work of consumers as they express themselves, their opinions and generate social solidarity with others in commercial digital media while, at the same time, adding economic value to those sites>>. This figure is an analogy showing that digitally performed labor has a similar relationship to capitalism as domestic labor, as both types of labor are essential for creating economic value and can produce non-marketable products. Additionally, there are interconnections and interdependencies between the spheres of production and reproduction in digital capitalism, which involve the accumulation of capital and exploitation. Within her study, Jarrett further examines the methods via which unpaid labor performed by social media users becomes integrated into oppressive economic systems.

Lastly, \citet{Bhattacharya2017} argues that the traditional employment model, which includes job security and full-time labor with one employer, is a historical exception that was mostly experienced by white workers in advanced countries during the Fordist era. Women, migrants, non-whites, and Global South workers have long had to work precariously since they were excluded from better occupations. From this perspective, precarious employment has consistently been the norm, and platform labor may simply assist these already vulnerable individuals by providing them with the opportunity for additional income and reducing their risks.

\subsection{Surveillance and control}
Feminist theories that deal with surveillance and control over labor often intersect with critical theories of technology, labor studies, and feminist critiques of capitalism. These theories analyze how surveillance practices are gendered and how they are used to control, exploit, and discipline labor, particularly in the context of neoliberal economies and digital technologies.
Although surveillance has become more prevalent due to advancements in technology and increased monitoring since the 1980s, its origins can be traced back to historical practices of state control, such as slavery, the regulation of women's reproductive rights, the monitoring of sexuality, and the scrutiny of impoverished individuals \citep{DUBROFSKY2015}.
The impact of emerging technologies on surveillance has been extensively studied, particularly in relation to concerns about privacy, which is being particularly impacted by the advent of new database technologies, advanced information storage systems, and modern communication technologies. However, privacy, although being an important perspective to consider when discussing surveillance, is restricted in its applicability due to the unequal distribution of this privilege. \citet{Hall2015} argues that a feminist approach to surveillance studies should aim to redirect the focus of critical surveillance studies from issues of privacy, security, and efficiency towards addressing the ethical challenge of addressing new forms of discrimination that occur in relation to categories of privilege, access, and risk.
Moreover, feminist analyses emphasize how surveillance is integral to numerous foundational structural systems, which foster marginalization and persist as established institutions. For instance, expanding on bell hooks's concept of ``white supremacist capitalist patriarchy'' \citep{hooks2000}, \citet{Smith2015} propose the term ``white supremacist capitalist hetero-patriarchal surveillance'' to describe the utilization of surveillance methods and technologies to establish and uphold whiteness, able-bodiedness, capitalism, and heterosexuality. All practices that are integral to the establishment and maintenance of the modern state.

The implementation of surveillance practices in the information economy has been altering power dynamics, potentially tilting the balance of power away from nation-states and towards large corporations that embrace the principles of \textit{surveillance capitalism}. \citet{Zuboff2019b} observes that surveillance capitalism goes beyond the traditional boundaries of private companies, acquiring not just surveillance assets and capital, but also rights, and functioning without effective processes of consent. Zuboff's theory defines surveillance capitalism as a novel market structure and a distinct method of capitalist accumulation. She describes it as a <<radically disembedded and extractive variant of information capitalism>> \citep{Zuboff2014}, which involves turning reality into behavioral data for analysis and sale. This is achieved by exploiting users of digital platforms who are essentially an unpaid workforce.

The emergence of ``epistemic inequality'', which refers to unequal access to learning due to private commercial mechanisms of information capture, production, analysis, and sales, indicates a shift in power from the ownership of means of production, which characterized 20th century politics, to the ownership of the production of meaning.
Surveillance capitalists exploit the increasing gap in knowledge distribution in order to generate profits. They exert control over the economy, society, and our lives without facing consequences, posing a threat not only to personal privacy but also to democracy. Moreover, <<unequal knowledge about us produces unequal power over us, and so epistemic inequality widens to include the distance between what we can do and what can be done to us>> \citep{Zuboff2020}.
The current privacy and antitrust rules are crucial but they alone will not be sufficient to address the emerging issues of reversing epistemic inequality. Rather, there is a need for a framework of epistemic rights that is legally protected and accountable to democratic governance.

Feminist scholars examine how surveillance capitalism disproportionately monitors and regulates the bodies, actions, and work of women, resulting in a form of surveillance that is influenced by gender \cite{Rizzo2023}. Similarly, it fosters many types of oppression by affecting marginalized categories based on race, class, and other identities in an intersectional manner \citep{Magnet2011}. In addition, as previously mentioned, feminist studies frequently examine the surveillance and exploitation of digital labor, specifically in the context of digital platform work and other precarious or informal employment, which has a substantial effect on women and marginalized communities. Regarding technology-enabled surveillance in the workplace, \citet{Chowdhary2023} has noted that the use of data-driven technologies has decreased the associated costs, thereby raising the potential for workplace surveillance and essentially removing any limitations.
Remarkably, algorithmic management utilizes surveillance, ratings, and rankings to exert control over workers, thereby converting market feedback into a system of disciplinary control \citep{Wang2024}.

\subsection{Blurring boundaries of employment}
As technology and society evolve, the conventional boundaries and structures of labor and employment are undergoing changes, as already underlined in \autoref{sec:techWork}.
The distinctions between self-employment, independent employment, dependent gainful employment, and informal employment are progressively eroding and becoming increasingly blurred. Similar observations may be made regarding productive and reproductive labor, the distinction between public and private domains, and the gender-based classifications that influence our perception of labor. This process of labor hybridization and fragmentation is also leading to the proliferation of phenomena such as the simultaneous engagement in self-employment and dependent employment, as well as frequent shifts between different types and sources of employment. Furthermore, the traditional boundaries between paid labor, leisure time, and care work are being blurred, as values is now generated also through activities such as internet browsing, mobile app usage, and the production of data traces. In addition, digital platforms establish the necessary structural framework for a global distribution, decentralization, and (a)synchronous execution of tasks, with a significant impact on work organization, the collective arrangement of workers, and the subsequent dissolution of both national and business organizational boundaries \citep{Huws2014,Rani2022,Schurmann2023}.

\citet{Schurmann2023} suggest reconsidering the metaphor of the cyborg, which represents a hybrid of a machine and organism existing in both social reality and fiction \citep{Haraway1985}, to deeper examine hybrid labor in the context of digital economy.
Haraway's notion of the cyborg is situated within a technologically saturated society, where distinctions between humans and machines, nature and technology, and other binary that shape our knowledge are eradicated. Within this utopia, technology has the capacity to bring about social change, specifically by dismantling the basic structure of hierarchical gender relations.
Indeed, the cyborg, as a figure of hybridity and dissolution of boundaries, symbolizes efficaciously the structural ambivalence of digital labor.
Furthermore, it is appropriate to address the overlapping of contradictorily structured dimensions of experience, such as self-determination and heteronomy, autonomy and control in the work process, the simultaneous presence of subordination and agency, as well as the generation of surplus and use value, while also considering the integration within societal hierarchies of domination \citep{Schurmann2023}.
In conclusion, Schürmann and Trenkmann observe that human participation in algorithm-supported processes of capital utilisation is affected by a process of double invisibilization: on one hand, it may seem like a voluntary engagement, but on the other hand, it actually leads to a loss of political agency. This follow a perception of technology that is based on two forms of negations: the first is the refusal to acknowledge the dominant character of technology, which is rooted in its monopolised ownership structure that grants exclusive access to information but remains hidden; the second is the denial of the necessity of care, as the integration of emotional care work into digital networks obscures the physical dependence that everyone has on care. Schürmann and Trenkmann contend that the cyborg effectively embodies this ambivalent nature; additionally, considering the convergence of subjectivity and technology, a collective identification as cyborgs presents the opportunity to overcome the existing divisions within the working class \cite{Schurmann2023,Bhattacharya2017}.


\subsection{Persistence of colonial lines}

According to research that combines feminist, Marxist, and postcolonial studies, platform economy functions by exploiting social constructs of race and gender, along with other intersectional dimensions. Racialized and gendered minorities are often seen as inexpensive labor due to the perception that they are easier to recruit and exploit. This perception is reinforced by beliefs about their low living standards and cultural inclination towards unpaid work, as well as by societal dynamics that marginalize and disempower them. This argument establishes a link between the colonial history and the mechanisms of global outsourcing.

Recent academic investigation has brought out the colonial facets of AI research \citep{Klein2024,Ricaurte2019}. 
Michael Kwet is one of the primary authors who thoroughly addressed the topic. He defines ``digital colonialism'' \cite{Kwet2019} as a systematic form of control exerted through the centralized ownership and regulation of the three fundamental components of the digital ecosystem: software, hardware, and network connectivity. This grants the United States (US) significant political, economic, and social influence. Therefore, GAFAM (Google/ Alphabet, Amazon, Facebook, Apple, and Microsoft) and other prominent corporations, along with state intelligence organizations, have emerged as the dominant forces in the international community, akin to imperialists. Foreign monopoly powers, with the US in the lead, are establishing infrastructure in the Global South with a design that serve their own interests, allowing them to exert economic and cultural control while enforcing privatized forms of governance. By doing so, they undercut local development, exert market dominance, and collect cash from the Global South by leveraging their authority derived primarily from the structural domination of digital architecture, resulting in broader forms of imperial control and colonial assimilation. 
Another important point to stree is that the workforce of the Global North is also vulnerable to the encroachment AI \citep*{Klein2024}. Indeed, marginalized groups within the Global North are also disproportionately subjected to digital surveillance, exploitation in the digital economy, algorithmic bias, and commodification via content appropriation. 

A similar argument has been put forward by Lelia Marie Hampton \cite{Hampton2023}. She conceptualized ``techno-racial capitalism'' as a theoretical framework, rooted in Black feminist Marxism, decolonial studies, and critical data studies, that explores how modern technology, specifically AI and machine learning, reinforces and intensifies the mechanisms of racial capitalism (i.e., a system in which socioeconomic value is extracted from racialized communities, often without any form of reciprocation, for capital accumulation). Just like Kwet, she argues that racialised groups are exploited by global digital industries in terms of data, labor, and resource extraction without providing them with any reciprocal benefits.

For the sake of completeness, US is not the only actor in the race for control over digital technologies. Currently, there are three prominent digital powers that hold significant influence, namely the US, China, and the European Union. These contemporary ``digital empires'' are the top leaders in technology, economy, and regulation, each possessing the aspiration and ability to influence the global digital structure according to their own interests and values. Each jurisdiction possesses a distinct perspective on the digital economy, which is evident in the regulatory frameworks they have implemented. The United States has spearheaded a predominantly market-oriented approach, China has adopted a state-driven model, and the European Union has embraced a right-based model \citep{Bradford2023}. 

Recently, there have been objections to the claim that the digital economy is structured based on colonial dynamics, and this criticism has led to further development of the subject. According, to \citet{Casilli2017}, theoretical comparisons between digital labor and slavery, imperialism, and colonization overlook the distinct historical nature of modern global inequities and do not move beyond vague, abstract analogies. Rather, he argues that the concept of ``coloniality'' provides a more accurate understanding of the social exclusion and exploitation that occur in both Western and non-Western countries.
In fact, while colonialism refers to the political and economic control of an empire over a colony, coloniality refers to enduring power dynamics that originated from colonialism and continue to shape culture, labor, relationships, and knowledge production, extending beyond the boundaries of colonial administrations \citep{Maldonado2007}.
In the digital economy, coloniality is not merely a metaphor, but rather a distinct and visible attribute that manifests in various ways. One such manifestation is its universal dynamic of marginalization within the workforce, where variations of intersectional categories operate to normalize the exploitation of underpaid or unpaid users. Moreover, economic structures establish ``colonial lines of demarcation'' that create divisions between human and nonhuman entities, as well as between elites and subalterns, and formal and invisible labor.

Coloniality serves as a powerful theoretical framework that exposes the underlying assumptions of dominant discourses and aims to achieve liberation for all marginalized identities in the workplace. \citet{Casilli2017} asserts that coloniality may be used as an analytical framework to identify and acknowledge the invisible digital workers and unrecognized tasks that are characteristic of the digital economy. This framework aims to facilitate a ``digital decolonial turn'', which involves <<making visible the invisible and about analyzing the mechanisms that produce such invisibility or distorted visibility in light of a large stock of ideas that must necessarily include the critical reflections of the ``invisible'' people themselves>> \citep{Maldonado2007}.
This objective also alignes with the seventh intersectional feminist principle proposed by \citet{Klein2024} , which states that: <<The work of data science, like all work in the world, is the work of many hands. Data feminism makes this labor visible so that it can be recognized and valued>>.

\section{Conclusion}
\label{sec:conclusion}
This paper has contributed to the understanding of digital platform work through feminist lens, offering a novel framework of four key dimensions (namely, precarity and economic exploitation, surveillance and control, blurring boundaries of employment, and persistence of colonial lines) to analyze how digital labor reinforces and reconfigures existing inequalities.
By situating these issues within broader historical and theoretical frameworks, we have underscored the ambivalent nature of automation, which result in its potential to empower workers being constrained by the entrenched power asymmetries it often amplifies.
This trend embodies an essential aspect of technology in general, and is especially prominent in recent advances driven by AI that also fuel digital platforms. 
The feminist dimensions we analyzed underscore the critical need to address the systemic inequities that shape digital labor markets, particularly through intersectional approaches that foreground marginalized voices and experiences, while striving to dismantle disparities within socio-economic power structures. 

The results indicate that although digital platforms may position themselves as facilitators of new employment opportunities that offer greater flexibility and accessibility relative to conventional job markets, they concurrently intensify economic exploitation, heighten surveillance, and blur the lines of formal employment, ultimately undermining worker rights and autonomy. Furthermore, digital platform labor, akin to domestic labor, often serves as an engine for the extraction of unpaid labor by transforming human activity into profit-generating labor. Additionally, it transmutes life itself into data for economic exploitation, thereby sustaining an increasingly oppressive loop.

Future research should expand on these dimensions by integrating participatory methodologies that actively engage digital workers in reimagining platform economies. One promising avenue is the exploration of digital platform cooperativism, which conveys a sustainable alternative to digital capitalism. This model proposes equitable and democratic labor practices founded on principles of decentralization, democratic co-ownership, and equitable value distribution \cite{Scholz2016a,Scholz2016b,Papadimitropoulos2021}.
Finally, further efforts are necessary to collect high-quality data regarding algorithmic management and working conditions on digital platforms. This will facilitate more accurate and rigorous empirical analysis, which is currently hindered by the difficulties associated with accessing data from privately owned digital platforms.

\bibliographystyle{apacite}
\bibliography{biblio}

\begin{thebibliography}{}

\bibitem [\protect \citeauthoryear {%
Agenjo-Calder{\'o}n%
\ \BBA {} G{\'a}lvez-Mu\~{n}oz%
}{%
Agenjo-Calder{\'o}n%
\ \BBA {} G{\'a}lvez-Mu\~{n}oz%
}{%
{\protect \APACyear {2019}}%
}]{%
Agenjo2019}
\APACinsertmetastar {%
Agenjo2019}%
\begin{APACrefauthors}%
Agenjo-Calder{\'o}n, A.%
\BCBT {}\ \BBA {} G{\'a}lvez-Mu\~{n}oz, L.%
\end{APACrefauthors}%
\unskip\
\newblock
\APACrefYearMonthDay{2019}{}{}.
\newblock
{\BBOQ}\APACrefatitle {Feminist Economics: Theoretical and Political Dimensions} {Feminist economics: Theoretical and political dimensions}.{\BBCQ}
\newblock
\APACjournalVolNumPages{The American Journal of Economics and Sociology}{78}{1}{137-166}.
\newblock
\begin{APACrefDOI} \doi{https://doi.org/10.1111/ajes.12264} \end{APACrefDOI}
\PrintBackRefs{\CurrentBib}

\bibitem [\protect \citeauthoryear {%
Altenried%
}{%
Altenried%
}{%
{\protect \APACyear {2022}}%
}]{%
altenried2022platforms}
\APACinsertmetastar {%
altenried2022platforms}%
\begin{APACrefauthors}%
Altenried, M.%
\end{APACrefauthors}%
\unskip\
\newblock
\APACrefYearMonthDay{2022}{}{}.
\newblock
{\BBOQ}\APACrefatitle {Platforms, labour, and mobility: migration and the gig economy} {Platforms, labour, and mobility: migration and the gig economy}.{\BBCQ}
\newblock
\BIn{} \APACrefbtitle {The Routledge Handbook of the Gig Economy} {The routledge handbook of the gig economy}\ (\BPGS\ 180--189).
\newblock
\APACaddressPublisher{}{Routledge}.
\PrintBackRefs{\CurrentBib}

\bibitem [\protect \citeauthoryear {%
Aronowitz%
\ \BBA {} Cutler%
}{%
Aronowitz%
\ \BBA {} Cutler%
}{%
{\protect \APACyear {1998}}%
}]{%
Aronowitz1998}
\APACinsertmetastar {%
Aronowitz1998}%
\begin{APACrefauthors}%
Aronowitz, S.%
\BCBT {}\ \BBA {} Cutler, J.%
\end{APACrefauthors}%
\ (\BEDS).
\unskip\
\newblock
\APACrefYear{1998}.
\newblock
\APACrefbtitle {{Post-Work}} {{Post-Work}}.
\newblock
\APACaddressPublisher{London, England}{Routledge}.
\PrintBackRefs{\CurrentBib}

\bibitem [\protect \citeauthoryear {%
Bengoa%
}{%
Bengoa%
}{%
{\protect \APACyear {2014}}%
}]{%
Carrasco2014}
\APACinsertmetastar {%
Carrasco2014}%
\begin{APACrefauthors}%
Bengoa, C\BPBI C.%
\end{APACrefauthors}%
\unskip\
\newblock
\APACrefYear{2014}.
\newblock
\APACrefbtitle {Con voz propia: La econom{\'\i}a feminista como apuesta te{\'o}rica y pol{\'\i}tica} {Con voz propia: La econom{\'\i}a feminista como apuesta te{\'o}rica y pol{\'\i}tica}.
\newblock
\APACaddressPublisher{}{La Oveja Roja}.
\PrintBackRefs{\CurrentBib}

\bibitem [\protect \citeauthoryear {%
Berardi%
}{%
Berardi%
}{%
{\protect \APACyear {2009}}%
}]{%
Berardi2009}
\APACinsertmetastar {%
Berardi2009}%
\begin{APACrefauthors}%
Berardi, F\BPBI B.%
\end{APACrefauthors}%
\unskip\
\newblock
\APACrefYear{2009}.
\newblock
\APACrefbtitle {The soul at work} {The soul at work}.
\newblock
\APACaddressPublisher{Brooklyn, NY}{Semiotext (E)}.
\PrintBackRefs{\CurrentBib}

\bibitem [\protect \citeauthoryear {%
Berardi%
}{%
Berardi%
}{%
{\protect \APACyear {2018}}%
}]{%
Bifo2018}
\APACinsertmetastar {%
Bifo2018}%
\begin{APACrefauthors}%
Berardi, F\BPBI B.%
\end{APACrefauthors}%
\unskip\
\newblock
\APACrefYear{2018}.
\newblock
\APACrefbtitle {Futurabilità} {Futurabilità}.
\newblock
\APACaddressPublisher{}{Nero Editions}.
\PrintBackRefs{\CurrentBib}

\bibitem [\protect \citeauthoryear {%
Bhattacharya%
}{%
Bhattacharya%
}{%
{\protect \APACyear {2017}}%
}]{%
Bhattacharya2017}
\APACinsertmetastar {%
Bhattacharya2017}%
\begin{APACrefauthors}%
Bhattacharya, T.%
\end{APACrefauthors}%
\unskip\
\newblock
\APACrefYear{2017}.
\newblock
\APACrefbtitle {Social Reproduction Theory: Remapping Class, Recentering Oppression} {Social reproduction theory: Remapping class, recentering oppression}.
\newblock
\APACaddressPublisher{}{Pluto Press}.
\newblock
\begin{APACrefURL} [{2024-08-19}]\url{http://www.jstor.org/stable/j.ctt1vz494j} \end{APACrefURL}
\PrintBackRefs{\CurrentBib}

\bibitem [\protect \citeauthoryear {%
Bookchin%
}{%
Bookchin%
}{%
{\protect \APACyear {1986}}%
}]{%
Bookchin1986}
\APACinsertmetastar {%
Bookchin1986}%
\begin{APACrefauthors}%
Bookchin, M.%
\end{APACrefauthors}%
\unskip\
\newblock
\APACrefYear{1986}.
\newblock
\APACrefbtitle {Post-scarcity Anarchism} {Post-scarcity anarchism}.
\newblock
\APACaddressPublisher{Montr{\'e}al, QC, Canada}{Black Rose Books}.
\PrintBackRefs{\CurrentBib}

\bibitem [\protect \citeauthoryear {%
Boscagli%
, Casarino%
, Colilli%
, Emory%
\BCBL {}\ \BBA {} Turits%
}{%
Boscagli%
\ \protect \BOthers {.}}{%
{\protect \APACyear {1996}}%
}]{%
Virno1996}
\APACinsertmetastar {%
Virno1996}%
\begin{APACrefauthors}%
Boscagli, M.%
, Casarino, C.%
, Colilli, P.%
, Emory, E.%
\BCBL {}\ \BBA {} Turits, M.%
\end{APACrefauthors}%
\unskip\
\newblock
\APACrefYear{1996}.
\newblock
\APACrefbtitle {Radical Thought in Italy: A Potential Politics} {Radical thought in italy: A potential politics}\ (\PrintOrdinal{NED - New edition}\ \BEd, \BVOL~7).
\newblock
\APACaddressPublisher{}{University of Minnesota Press}.
\newblock
\begin{APACrefURL} [{2025-01-06}]\url{http://www.jstor.org/stable/10.5749/j.ctttssjm} \end{APACrefURL}
\PrintBackRefs{\CurrentBib}

\bibitem [\protect \citeauthoryear {%
Bradford%
}{%
Bradford%
}{%
{\protect \APACyear {2023}}%
}]{%
Bradford2023}
\APACinsertmetastar {%
Bradford2023}%
\begin{APACrefauthors}%
Bradford, A.%
\end{APACrefauthors}%
\unskip\
\newblock
\APACrefYear{2023}.
\newblock
\APACrefbtitle {Digital empires} {Digital empires}.
\newblock
\APACaddressPublisher{New York, NY}{Oxford University Press}.
\PrintBackRefs{\CurrentBib}

\bibitem [\protect \citeauthoryear {%
Breman%
}{%
Breman%
}{%
{\protect \APACyear {2013}}%
}]{%
Breman2013}
\APACinsertmetastar {%
Breman2013}%
\begin{APACrefauthors}%
Breman, J.%
\end{APACrefauthors}%
\unskip\
\newblock
\APACrefYearMonthDay{2013}{}{}.
\newblock
{\BBOQ}\APACrefatitle {A Bogus Concept?} {A bogus concept?}{\BBCQ}
\newblock
\APACjournalVolNumPages{New Left Review}{}{}{130–138}.
\newblock
\begin{APACrefURL} \url{https://newleftreview.org/issues/ii84/articles/jan-breman-a-bogus-concept} \end{APACrefURL}
\PrintBackRefs{\CurrentBib}

\bibitem [\protect \citeauthoryear {%
Casilli%
}{%
Casilli%
}{%
{\protect \APACyear {2017}}%
}]{%
Casilli2017}
\APACinsertmetastar {%
Casilli2017}%
\begin{APACrefauthors}%
Casilli, A\BPBI A.%
\end{APACrefauthors}%
\unskip\
\newblock
\APACrefYearMonthDay{2017}{}{}.
\newblock
{\BBOQ}\APACrefatitle {Digital labor studies go global: Toward a digital decolonial turn} {Digital labor studies go global: Toward a digital decolonial turn}.{\BBCQ}
\newblock
\APACjournalVolNumPages{International Journal of Communication}{11}{}{21}.
\PrintBackRefs{\CurrentBib}

\bibitem [\protect \citeauthoryear {%
Chowdhary%
\ \protect \BOthers {.}}{%
Chowdhary%
\ \protect \BOthers {.}}{%
{\protect \APACyear {2023}}%
}]{%
Chowdhary2023}
\APACinsertmetastar {%
Chowdhary2023}%
\begin{APACrefauthors}%
Chowdhary, S.%
, Kawakami, A.%
, Gray, M\BPBI L.%
, Suh, J.%
, Olteanu, A.%
\BCBL {}\ \BBA {} Saha, K.%
\end{APACrefauthors}%
\unskip\
\newblock
\APACrefYearMonthDay{2023}{}{}.
\newblock
{\BBOQ}\APACrefatitle {Can Workers Meaningfully Consent to Workplace Wellbeing Technologies?} {Can workers meaningfully consent to workplace wellbeing technologies?}{\BBCQ}
\newblock
\BIn{} \APACrefbtitle {Proceedings of the 2023 ACM Conference on Fairness, Accountability, and Transparency} {Proceedings of the 2023 acm conference on fairness, accountability, and transparency}\ (\BPG~569–582).
\newblock
\APACaddressPublisher{New York, NY, USA}{Association for Computing Machinery}.
\newblock
\begin{APACrefURL} \url{https://doi.org/10.1145/3593013.3594023} \end{APACrefURL}
\newblock
\begin{APACrefDOI} \doi{10.1145/3593013.3594023} \end{APACrefDOI}
\PrintBackRefs{\CurrentBib}

\bibitem [\protect \citeauthoryear {%
Cirillo%
, Guarascio%
\BCBL {}\ \BBA {} Parolin%
}{%
Cirillo%
\ \protect \BOthers {.}}{%
{\protect \APACyear {2023}}%
}]{%
Cirillo2023}
\APACinsertmetastar {%
Cirillo2023}%
\begin{APACrefauthors}%
Cirillo, V.%
, Guarascio, D.%
\BCBL {}\ \BBA {} Parolin, Z.%
\end{APACrefauthors}%
\unskip\
\newblock
\APACrefYearMonthDay{2023}{{\APACmonth{06}}}{}.
\newblock
{\BBOQ}\APACrefatitle {Platform work and economic insecurity in Italy} {Platform work and economic insecurity in italy}.{\BBCQ}
\newblock
\APACjournalVolNumPages{Structural Change and Economic Dynamics}{65}{}{126–138}.
\newblock
\begin{APACrefDOI} \doi{10.1016/j.strueco.2023.02.011} \end{APACrefDOI}
\PrintBackRefs{\CurrentBib}

\bibitem [\protect \citeauthoryear {%
Couldry%
\ \BBA {} Mejias%
}{%
Couldry%
\ \BBA {} Mejias%
}{%
{\protect \APACyear {2019}}%
}]{%
Couldry2019}
\APACinsertmetastar {%
Couldry2019}%
\begin{APACrefauthors}%
Couldry, N.%
\BCBT {}\ \BBA {} Mejias, U\BPBI A.%
\end{APACrefauthors}%
\unskip\
\newblock
\APACrefYear{2019}.
\newblock
\APACrefbtitle {The costs of connection: How Data Is Colonizing Human Life and Appropriating It for Capitalism} {The costs of connection: How data is colonizing human life and appropriating it for capitalism}.
\newblock
\APACaddressPublisher{Palo Alto, CA}{Stanford University Press}.
\PrintBackRefs{\CurrentBib}

\bibitem [\protect \citeauthoryear {%
Coveri%
, Cozza%
\BCBL {}\ \BBA {} Guarascio%
}{%
Coveri%
\ \protect \BOthers {.}}{%
{\protect \APACyear {2022}}%
}]{%
Coveri2022}
\APACinsertmetastar {%
Coveri2022}%
\begin{APACrefauthors}%
Coveri, A.%
, Cozza, C.%
\BCBL {}\ \BBA {} Guarascio, D.%
\end{APACrefauthors}%
\unskip\
\newblock
\APACrefYearMonthDay{2022}{{\APACmonth{10}}}{}.
\newblock
{\BBOQ}\APACrefatitle {Monopoly Capital in the time of digital platforms: a radical approach to the Amazon case} {Monopoly capital in the time of digital platforms: a radical approach to the amazon case}.{\BBCQ}
\newblock
\APACjournalVolNumPages{Cambridge Journal of Economics}{46}{6}{1341–1367}.
\newblock
\begin{APACrefDOI} \doi{10.1093/cje/beac044} \end{APACrefDOI}
\PrintBackRefs{\CurrentBib}

\bibitem [\protect \citeauthoryear {%
Cukier%
}{%
Cukier%
}{%
{\protect \APACyear {2018}}%
}]{%
Cukier2018}
\APACinsertmetastar {%
Cukier2018}%
\begin{APACrefauthors}%
Cukier, A.%
\end{APACrefauthors}%
\unskip\
\newblock
\APACrefYear{2018}.
\newblock
\APACrefbtitle {Qu'est-ce que le travail ?} {Qu'est-ce que le travail ?}
\newblock
\APACaddressPublisher{}{Vrin - Chemins Philosophiques}.
\PrintBackRefs{\CurrentBib}

\bibitem [\protect \citeauthoryear {%
Dalla~Costa%
\ \BBA {} James%
}{%
Dalla~Costa%
\ \BBA {} James%
}{%
{\protect \APACyear {2017}}%
}]{%
DallaCosta2017}
\APACinsertmetastar {%
DallaCosta2017}%
\begin{APACrefauthors}%
Dalla~Costa, M.%
\BCBT {}\ \BBA {} James, S.%
\end{APACrefauthors}%
\unskip\
\newblock
\APACrefYearMonthDay{2017}{}{}.
\newblock
{\BBOQ}\APACrefatitle {The Power of Women and the Subversion of the Community} {The power of women and the subversion of the community}.{\BBCQ}
\newblock
\APACjournalVolNumPages{Class: The Anthology}{}{}{79--86}.
\PrintBackRefs{\CurrentBib}

\bibitem [\protect \citeauthoryear {%
Dubrofsky%
\ \BBA {} Magnet%
}{%
Dubrofsky%
\ \BBA {} Magnet%
}{%
{\protect \APACyear {2015}}%
}]{%
DUBROFSKY2015}
\APACinsertmetastar {%
DUBROFSKY2015}%
\begin{APACrefauthors}%
Dubrofsky, R\BPBI E.%
\BCBT {}\ \BBA {} Magnet, S\BPBI A.%
\end{APACrefauthors}%
\unskip\
\newblock
\APACrefYearMonthDay{2015}{{\APACmonth{05}}}{}.
\newblock
{\BBOQ}\APACrefatitle {Feminist Surveillance Studies: Critical Interventions} {Feminist surveillance studies: Critical interventions}.{\BBCQ}
\newblock
\BIn{} \APACrefbtitle {Feminist Surveillance Studies} {Feminist surveillance studies}\ (\BPG~1–18).
\newblock
\APACaddressPublisher{}{Duke University Press}.
\newblock
\begin{APACrefDOI} \doi{10.2307/j.ctv1198x2b.5} \end{APACrefDOI}
\PrintBackRefs{\CurrentBib}

\bibitem [\protect \citeauthoryear {%
Elwood%
}{%
Elwood%
}{%
{\protect \APACyear {2020}}%
}]{%
Elwood2020}
\APACinsertmetastar {%
Elwood2020}%
\begin{APACrefauthors}%
Elwood, S.%
\end{APACrefauthors}%
\unskip\
\newblock
\APACrefYearMonthDay{2020}{{\APACmonth{01}}}{}.
\newblock
{\BBOQ}\APACrefatitle {Digital geographies, feminist relationality, Black and queer code studies: Thriving otherwise} {Digital geographies, feminist relationality, black and queer code studies: Thriving otherwise}.{\BBCQ}
\newblock
\APACjournalVolNumPages{Progress in Human Geography}{45}{2}{209–228}.
\newblock
\begin{APACrefDOI} \doi{10.1177/0309132519899733} \end{APACrefDOI}
\PrintBackRefs{\CurrentBib}

\bibitem [\protect \citeauthoryear {%
Fayard%
}{%
Fayard%
}{%
{\protect \APACyear {2021}}%
}]{%
Fayard2021}
\APACinsertmetastar {%
Fayard2021}%
\begin{APACrefauthors}%
Fayard, A\BHBI L.%
\end{APACrefauthors}%
\unskip\
\newblock
\APACrefYearMonthDay{2021}{}{}.
\newblock
{\BBOQ}\APACrefatitle {Notes on the Meaning of Work: Labor, Work, and Action in the 21st Century} {Notes on the meaning of work: Labor, work, and action in the 21st century}.{\BBCQ}
\newblock
\APACjournalVolNumPages{Journal of Management Inquiry}{30}{2}{207-220}.
\newblock
\begin{APACrefDOI} \doi{10.1177/1056492619841705} \end{APACrefDOI}
\PrintBackRefs{\CurrentBib}

\bibitem [\protect \citeauthoryear {%
Federici%
\ \BBA {} Fortunati%
}{%
Federici%
\ \BBA {} Fortunati%
}{%
{\protect \APACyear {1984}}%
}]{%
Federici1984}
\APACinsertmetastar {%
Federici1984}%
\begin{APACrefauthors}%
Federici, S.%
\BCBT {}\ \BBA {} Fortunati, L.%
\end{APACrefauthors}%
\unskip\
\newblock
\APACrefYear{1984}.
\newblock
\APACrefbtitle {Il grande Calibano: storia del corpo sociale ribelle nella prima fase del capitale} {Il grande calibano: storia del corpo sociale ribelle nella prima fase del capitale}.
\newblock
\APACaddressPublisher{}{F. Angeli}.
\newblock
\begin{APACrefURL} \url{https://books.google.it/books?id=LQbfPAAACAAJ} \end{APACrefURL}
\PrintBackRefs{\CurrentBib}

\bibitem [\protect \citeauthoryear {%
Fortunati%
}{%
Fortunati%
}{%
{\protect \APACyear {1995}}%
}]{%
Fortunati1995}
\APACinsertmetastar {%
Fortunati1995}%
\begin{APACrefauthors}%
Fortunati, L.%
\end{APACrefauthors}%
\unskip\
\newblock
\APACrefYearMonthDay{1995}{}{}.
\newblock
{\BBOQ}\APACrefatitle {The Arcane of Reproduction: Housework, Prostitution, Labor and Capital, translated by H} {The arcane of reproduction: Housework, prostitution, labor and capital, translated by h}.{\BBCQ}
\newblock
\APACjournalVolNumPages{Creek. Brooklyn: Autonomedia}{}{}{}.
\PrintBackRefs{\CurrentBib}

\bibitem [\protect \citeauthoryear {%
Hall%
}{%
Hall%
}{%
{\protect \APACyear {2015}}%
}]{%
Hall2015}
\APACinsertmetastar {%
Hall2015}%
\begin{APACrefauthors}%
Hall, R.%
\end{APACrefauthors}%
\unskip\
\newblock
\APACrefYearMonthDay{2015}{}{}.
\newblock
{\BBOQ}\APACrefatitle {Terror and the Female Grotesque: Introducing Full-Body Scanners to U.S. Airports} {Terror and the female grotesque: Introducing full-body scanners to u.s. airports}.{\BBCQ}
\newblock
\BIn{} \APACrefbtitle {Feminist Surveillance Studies.} {Feminist surveillance studies.}
\newblock
\APACaddressPublisher{}{Duke University Press}.
\newblock
\begin{APACrefURL} \url{https://api.semanticscholar.org/CorpusID:131834206} \end{APACrefURL}
\PrintBackRefs{\CurrentBib}

\bibitem [\protect \citeauthoryear {%
Hampton%
}{%
Hampton%
}{%
{\protect \APACyear {2023}}%
}]{%
Hampton2023}
\APACinsertmetastar {%
Hampton2023}%
\begin{APACrefauthors}%
Hampton, L\BPBI M.%
\end{APACrefauthors}%
\unskip\
\newblock
\APACrefYearMonthDay{2023}{10}{}.
\newblock
{\BBOQ}\APACrefatitle {{119Techno-Racial Capitalism: A Decolonial Black Feminist Marxist Perspective}} {{119Techno-Racial Capitalism: A Decolonial Black Feminist Marxist Perspective}}.{\BBCQ}
\newblock
\BIn{} \APACrefbtitle {{Feminist AI: Critical Perspectives on Algorithms, Data, and Intelligent Machines}.} {{Feminist AI: Critical Perspectives on Algorithms, Data, and Intelligent Machines}.}
\newblock
\APACaddressPublisher{}{Oxford University Press}.
\newblock
\begin{APACrefDOI} \doi{10.1093/oso/9780192889898.003.0008} \end{APACrefDOI}
\PrintBackRefs{\CurrentBib}

\bibitem [\protect \citeauthoryear {%
Haraway%
}{%
Haraway%
}{%
{\protect \APACyear {1985}}%
}]{%
Haraway1985}
\APACinsertmetastar {%
Haraway1985}%
\begin{APACrefauthors}%
Haraway, D\BPBI J.%
\end{APACrefauthors}%
\unskip\
\newblock
\APACrefYearMonthDay{1985}{}{}.
\newblock
{\BBOQ}\APACrefatitle {A manifesto for Cyborgs: Science, technology, and socialist feminism in the 1980s} {A manifesto for cyborgs: Science, technology, and socialist feminism in the 1980s}.{\BBCQ}
\newblock
\APACjournalVolNumPages{Socialist Review 80}{15}{}{65-107}.
\PrintBackRefs{\CurrentBib}

\bibitem [\protect \citeauthoryear {%
Hardt%
\ \BBA {} Negri%
}{%
Hardt%
\ \BBA {} Negri%
}{%
{\protect \APACyear {2005}}%
}]{%
Hardt2005}
\APACinsertmetastar {%
Hardt2005}%
\begin{APACrefauthors}%
Hardt, M.%
\BCBT {}\ \BBA {} Negri, A.%
\end{APACrefauthors}%
\unskip\
\newblock
\APACrefYear{2005}.
\newblock
\APACrefbtitle {Multitude: War and democracy in the age of empire} {Multitude: War and democracy in the age of empire}.
\newblock
\APACaddressPublisher{}{Penguin}.
\PrintBackRefs{\CurrentBib}

\bibitem [\protect \citeauthoryear {%
Hegel%
}{%
Hegel%
}{%
{\protect \APACyear {1991}}%
}]{%
Hegel1991}
\APACinsertmetastar {%
Hegel1991}%
\begin{APACrefauthors}%
Hegel, G\BPBI W\BPBI F.%
\end{APACrefauthors}%
\unskip\
\newblock
\APACrefYear{1991}.
\newblock
\APACrefbtitle {Cambridge texts in the history of political thought: Hegel: Elements of the philosophy of right} {Cambridge texts in the history of political thought: Hegel: Elements of the philosophy of right}\ (A\BPBI W.~Wood, \BED{}).
\newblock
\APACaddressPublisher{Cambridge, England}{Cambridge University Press}.
\PrintBackRefs{\CurrentBib}

\bibitem [\protect \citeauthoryear {%
hooks%
}{%
hooks%
}{%
{\protect \APACyear {2000}}%
}]{%
hooks2000}
\APACinsertmetastar {%
hooks2000}%
\begin{APACrefauthors}%
hooks, b.%
\end{APACrefauthors}%
\unskip\
\newblock
\APACrefYear{2000}.
\newblock
\APACrefbtitle {Feminist theory: From margin to center} {Feminist theory: From margin to center}.
\newblock
\APACaddressPublisher{}{Pluto Press}.
\PrintBackRefs{\CurrentBib}

\bibitem [\protect \citeauthoryear {%
Huws%
}{%
Huws%
}{%
{\protect \APACyear {2014}}%
}]{%
Huws2014}
\APACinsertmetastar {%
Huws2014}%
\begin{APACrefauthors}%
Huws, U.%
\end{APACrefauthors}%
\unskip\
\newblock
\APACrefYear{2014}.
\newblock
\APACrefbtitle {Labor in the Global Digital Economy: The Cybertariat Comes of Age} {Labor in the global digital economy: The cybertariat comes of age}.
\newblock
\APACaddressPublisher{}{NYU Press}.
\newblock
\begin{APACrefURL} [{2024-08-09}]\url{http://www.jstor.org/stable/j.ctt1287j8b} \end{APACrefURL}
\PrintBackRefs{\CurrentBib}

\bibitem [\protect \citeauthoryear {%
Huws%
}{%
Huws%
}{%
{\protect \APACyear {2019}}%
}]{%
Huws2019}
\APACinsertmetastar {%
Huws2019}%
\begin{APACrefauthors}%
Huws, U.%
\end{APACrefauthors}%
\unskip\
\newblock
\APACrefYearMonthDay{2019}{}{}.
\newblock
{\BBOQ}\APACrefatitle {The Hassle of Housework: Digitalisation and the Commodification of Domestic Labour} {The hassle of housework: Digitalisation and the commodification of domestic labour}.{\BBCQ}
\newblock
\APACjournalVolNumPages{Feminist Review}{123}{1}{8-23}.
\newblock
\begin{APACrefDOI} \doi{10.1177/0141778919879725} \end{APACrefDOI}
\PrintBackRefs{\CurrentBib}

\bibitem [\protect \citeauthoryear {%
Huws%
\ \BBA {} Leys%
}{%
Huws%
\ \BBA {} Leys%
}{%
{\protect \APACyear {2003}}%
}]{%
Huws2003}
\APACinsertmetastar {%
Huws2003}%
\begin{APACrefauthors}%
Huws, U.%
\BCBT {}\ \BBA {} Leys, C.%
\end{APACrefauthors}%
\unskip\
\newblock
\APACrefYear{2003}.
\newblock
\APACrefbtitle {The Making of a Cybertariat: Virtual Work in a Real World} {The making of a cybertariat: Virtual work in a real world}.
\newblock
\APACaddressPublisher{USA}{Monthly Review Press}.
\PrintBackRefs{\CurrentBib}

\bibitem [\protect \citeauthoryear {%
Hymer%
}{%
Hymer%
}{%
{\protect \APACyear {1982}}%
}]{%
Hymer1982}
\APACinsertmetastar {%
Hymer1982}%
\begin{APACrefauthors}%
Hymer, S.%
\end{APACrefauthors}%
\unskip\
\newblock
\APACrefYearMonthDay{1982}{}{}.
\newblock
{\BBOQ}\APACrefatitle {The Multinational Corporation and the Law of Uneven Development} {The multinational corporation and the law of uneven development}.{\BBCQ}
\newblock
\BIn{} \APACrefbtitle {Introduction to the Sociology of ''Developing Societies'''} {Introduction to the sociology of ''developing societies'''}\ (\BPG~128–152).
\newblock
\APACaddressPublisher{}{Macmillan Education UK}.
\newblock
\begin{APACrefDOI} \doi{10.1007/978-1-349-16847-7_12} \end{APACrefDOI}
\PrintBackRefs{\CurrentBib}

\bibitem [\protect \citeauthoryear {%
{Into the Black Box}%
}{%
{Into the Black Box}%
}{%
{\protect \APACyear {2021}}%
}]{%
IntoTheBlackBox}
\APACinsertmetastar {%
IntoTheBlackBox}%
\begin{APACrefauthors}%
{Into the Black Box}.%
\end{APACrefauthors}%
\unskip\
\newblock
\APACrefYearMonthDay{2021}{}{}.
\newblock
{\BBOQ}\APACrefatitle {Labour and Technology in Capitalism 4.0. An Interview with Ursula Huws} {Labour and technology in capitalism 4.0. an interview with ursula huws}.{\BBCQ}
\newblock
\APACjournalVolNumPages{Soft Power}{}{}{164-178}.
\newblock
\begin{APACrefDOI} \doi{10.14718/SoftPower.2021.8.1.9} \end{APACrefDOI}
\PrintBackRefs{\CurrentBib}

\bibitem [\protect \citeauthoryear {%
Jarrett%
}{%
Jarrett%
}{%
{\protect \APACyear {2015}}%
}]{%
Jarrett2015}
\APACinsertmetastar {%
Jarrett2015}%
\begin{APACrefauthors}%
Jarrett, K.%
\end{APACrefauthors}%
\unskip\
\newblock
\APACrefYear{2015}.
\newblock
\APACrefbtitle {Feminism, Labour and Digital Media: The Digital Housewife} {Feminism, labour and digital media: The digital housewife}.
\newblock
\APACaddressPublisher{}{Routledge}.
\PrintBackRefs{\CurrentBib}

\bibitem [\protect \citeauthoryear {%
Keynes%
}{%
Keynes%
}{%
{\protect \APACyear {2010}}%
}]{%
Keynes2010}
\APACinsertmetastar {%
Keynes2010}%
\begin{APACrefauthors}%
Keynes, J\BPBI M.%
\end{APACrefauthors}%
\unskip\
\newblock
\APACrefYearMonthDay{2010}{}{}.
\newblock
{\BBOQ}\APACrefatitle {Economic Possibilities for Our Grandchildren} {Economic possibilities for our grandchildren}.{\BBCQ}
\newblock
\BIn{} \APACrefbtitle {Essays in Persuasion} {Essays in persuasion}\ (\BPGS\ 321--332).
\newblock
\APACaddressPublisher{London}{Palgrave Macmillan UK}.
\newblock
\begin{APACrefDOI} \doi{10.1007/978-1-349-59072-8_25} \end{APACrefDOI}
\PrintBackRefs{\CurrentBib}

\bibitem [\protect \citeauthoryear {%
Klein%
\ \BBA {} D'Ignazio%
}{%
Klein%
\ \BBA {} D'Ignazio%
}{%
{\protect \APACyear {2024}}%
}]{%
Klein2024}
\APACinsertmetastar {%
Klein2024}%
\begin{APACrefauthors}%
Klein, L.%
\BCBT {}\ \BBA {} D'Ignazio, C.%
\end{APACrefauthors}%
\unskip\
\newblock
\APACrefYearMonthDay{2024}{{\APACmonth{06}}}{}.
\newblock
{\BBOQ}\APACrefatitle {Data Feminism for {AI}} {Data feminism for {AI}}.{\BBCQ}
\newblock
\BIn{} \APACrefbtitle {The 2024 ACM Conference on Fairness, Accountability, and Transparency.} {The 2024 acm conference on fairness, accountability, and transparency.}
\newblock
\APACaddressPublisher{}{ACM}.
\newblock
\begin{APACrefDOI} \doi{10.1145/3630106.3658543} \end{APACrefDOI}
\PrintBackRefs{\CurrentBib}

\bibitem [\protect \citeauthoryear {%
Kwet%
}{%
Kwet%
}{%
{\protect \APACyear {2019}}%
}]{%
Kwet2019}
\APACinsertmetastar {%
Kwet2019}%
\begin{APACrefauthors}%
Kwet, M.%
\end{APACrefauthors}%
\unskip\
\newblock
\APACrefYearMonthDay{2019}{}{}.
\newblock
{\BBOQ}\APACrefatitle {Digital colonialism: US empire and the new imperialism in the Global South} {Digital colonialism: Us empire and the new imperialism in the global south}.{\BBCQ}
\newblock
\APACjournalVolNumPages{Race \& Class}{60}{4}{3-26}.
\newblock
\begin{APACrefDOI} \doi{10.1177/0306396818823172} \end{APACrefDOI}
\PrintBackRefs{\CurrentBib}

\bibitem [\protect \citeauthoryear {%
Kyrylych%
}{%
Kyrylych%
}{%
{\protect \APACyear {2013}}%
}]{%
Kyrylych2013}
\APACinsertmetastar {%
Kyrylych2013}%
\begin{APACrefauthors}%
Kyrylych, K.%
\end{APACrefauthors}%
\unskip\
\newblock
\APACrefYearMonthDay{2013}{}{}.
\newblock
{\BBOQ}\APACrefatitle {Problem of Uneven Economic Development of the World Economy: Essence and Causes} {Problem of uneven economic development of the world economy: Essence and causes}.{\BBCQ}
\newblock
\APACjournalVolNumPages{Intellectual Economics}{7}{3}{344–354}.
\newblock
\begin{APACrefDOI} \doi{10.13165/ie-13-7-3-06} \end{APACrefDOI}
\PrintBackRefs{\CurrentBib}

\bibitem [\protect \citeauthoryear {%
Lyn~Hoang%
\ \BBA {} Quan-Haase%
}{%
Lyn~Hoang%
\ \BBA {} Quan-Haase%
}{%
{\protect \APACyear {2020}}%
}]{%
Hoang2020}
\APACinsertmetastar {%
Hoang2020}%
\begin{APACrefauthors}%
Lyn~Hoang, G\BPBI B.%
\BCBT {}\ \BBA {} Quan-Haase, A.%
\end{APACrefauthors}%
\unskip\
\newblock
\APACrefYearMonthDay{2020}{}{}.
\newblock
{\BBOQ}\APACrefatitle {The winners and the losers of the platform economy: who participates?} {The winners and the losers of the platform economy: who participates?}{\BBCQ}
\newblock
\APACjournalVolNumPages{Information, Communication \& Society}{23}{5}{681--700}.
\newblock
\begin{APACrefDOI} \doi{10.1080/1369118X.2020.1720771} \end{APACrefDOI}
\PrintBackRefs{\CurrentBib}

\bibitem [\protect \citeauthoryear {%
Magnet%
}{%
Magnet%
}{%
{\protect \APACyear {2011}}%
}]{%
Magnet2011}
\APACinsertmetastar {%
Magnet2011}%
\begin{APACrefauthors}%
Magnet, S\BPBI A.%
\end{APACrefauthors}%
\unskip\
\newblock
\APACrefYear{2011}.
\newblock
\APACrefbtitle {When Biometrics Fail: Gender, Race, and the Technology of Identity} {When biometrics fail: Gender, race, and the technology of identity}.
\newblock
\APACaddressPublisher{}{Duke University Press}.
\newblock
\begin{APACrefURL} [{2024-08-19}]\url{http://www.jstor.org/stable/j.ctv12101s4} \end{APACrefURL}
\PrintBackRefs{\CurrentBib}

\bibitem [\protect \citeauthoryear {%
Maldonado-Torres%
}{%
Maldonado-Torres%
}{%
{\protect \APACyear {2007}}%
}]{%
Maldonado2007}
\APACinsertmetastar {%
Maldonado2007}%
\begin{APACrefauthors}%
Maldonado-Torres, N.%
\end{APACrefauthors}%
\unskip\
\newblock
\APACrefYearMonthDay{2007}{}{}.
\newblock
{\BBOQ}\APACrefatitle {On the coloniality of being} {On the coloniality of being}.{\BBCQ}
\newblock
\APACjournalVolNumPages{Cultural Studies}{21}{2-3}{240--270}.
\newblock
\begin{APACrefDOI} \doi{10.1080/09502380601162548} \end{APACrefDOI}
\PrintBackRefs{\CurrentBib}

\bibitem [\protect \citeauthoryear {%
Marcuse%
}{%
Marcuse%
}{%
{\protect \APACyear {1955}}%
}]{%
Marcuse1955}
\APACinsertmetastar {%
Marcuse1955}%
\begin{APACrefauthors}%
Marcuse, H.%
\end{APACrefauthors}%
\unskip\
\newblock
\APACrefYear{1955}.
\newblock
\APACrefbtitle {Eros And Civilization: a philosophical inquiry into Freud} {Eros and civilization: a philosophical inquiry into freud}.
\newblock
\APACaddressPublisher{}{Beacon Press}.
\PrintBackRefs{\CurrentBib}

\bibitem [\protect \citeauthoryear {%
Marcuse%
}{%
Marcuse%
}{%
{\protect \APACyear {1991}}%
}]{%
Marcuse1964}
\APACinsertmetastar {%
Marcuse1964}%
\begin{APACrefauthors}%
Marcuse, H.%
\end{APACrefauthors}%
\unskip\
\newblock
\APACrefYear{1991}.
\newblock
\APACrefbtitle {{One-Dimensional Man}} {{One-Dimensional Man}}\ (\PrintOrdinal{2}\ \BEd).
\newblock
\APACaddressPublisher{London, England}{Routledge}.
\PrintBackRefs{\CurrentBib}

\bibitem [\protect \citeauthoryear {%
Marx%
}{%
Marx%
}{%
{\protect \APACyear {1961}}%
}]{%
Marx1961}
\APACinsertmetastar {%
Marx1961}%
\begin{APACrefauthors}%
Marx, K.%
\end{APACrefauthors}%
\unskip\
\newblock
\APACrefYear{1961}.
\newblock
\APACrefbtitle {Capital} {Capital}.
\newblock
\APACaddressPublisher{}{Foreign Languages Publishing House, Moscow, 1st edn, 1867}.
\PrintBackRefs{\CurrentBib}

\bibitem [\protect \citeauthoryear {%
Marx%
}{%
Marx%
}{%
{\protect \APACyear {1973}}%
}]{%
Marx1973}
\APACinsertmetastar {%
Marx1973}%
\begin{APACrefauthors}%
Marx, K.%
\end{APACrefauthors}%
\unskip\
\newblock
\APACrefYear{1973}.
\newblock
\APACrefbtitle {Grundrisse} {Grundrisse}.
\newblock
\APACaddressPublisher{Harlow, England}{Penguin Books}.
\PrintBackRefs{\CurrentBib}

\bibitem [\protect \citeauthoryear {%
Mies%
, Office%
\BCBL {}\ \BBA {} Programme%
}{%
Mies%
\ \protect \BOthers {.}}{%
{\protect \APACyear {1982}}%
}]{%
Mies1982}
\APACinsertmetastar {%
Mies1982}%
\begin{APACrefauthors}%
Mies, M.%
, Office, I\BPBI L.%
\BCBL {}\ \BBA {} Programme, W\BPBI E.%
\end{APACrefauthors}%
\unskip\
\newblock
\APACrefYear{1982}.
\newblock
\APACrefbtitle {The Lace Makers of Narsapur: Indian Housewives Produce for the World Market} {The lace makers of narsapur: Indian housewives produce for the world market}.
\newblock
\APACaddressPublisher{}{Zed Press}.
\newblock
\begin{APACrefURL} \url{https://books.google.it/books?id=RTp1AAAAIAAJ} \end{APACrefURL}
\PrintBackRefs{\CurrentBib}

\bibitem [\protect \citeauthoryear {%
Muldoon%
\ \BBA {} Raekstad%
}{%
Muldoon%
\ \BBA {} Raekstad%
}{%
{\protect \APACyear {2023}}%
}]{%
Muldoon23}
\APACinsertmetastar {%
Muldoon23}%
\begin{APACrefauthors}%
Muldoon, J.%
\BCBT {}\ \BBA {} Raekstad, P.%
\end{APACrefauthors}%
\unskip\
\newblock
\APACrefYearMonthDay{2023}{}{}.
\newblock
{\BBOQ}\APACrefatitle {Algorithmic domination in the gig economy} {Algorithmic domination in the gig economy}.{\BBCQ}
\newblock
\APACjournalVolNumPages{European Journal of Political Theory}{22}{4}{587-607}.
\newblock
\begin{APACrefDOI} \doi{10.1177/14748851221082078} \end{APACrefDOI}
\PrintBackRefs{\CurrentBib}

\bibitem [\protect \citeauthoryear {%
Muldoon%
\ \BBA {} Wu%
}{%
Muldoon%
\ \BBA {} Wu%
}{%
{\protect \APACyear {2023}}%
}]{%
Muldoon2023Matrix}
\APACinsertmetastar {%
Muldoon2023Matrix}%
\begin{APACrefauthors}%
Muldoon, J.%
\BCBT {}\ \BBA {} Wu, B\BPBI A.%
\end{APACrefauthors}%
\unskip\
\newblock
\APACrefYearMonthDay{2023}{{\APACmonth{12}}}{}.
\newblock
{\BBOQ}\APACrefatitle {Artificial Intelligence in the Colonial Matrix of Power} {Artificial intelligence in the colonial matrix of power}.{\BBCQ}
\newblock
\APACjournalVolNumPages{Philosophy \& Technology}{36}{4}{}.
\newblock
\begin{APACrefURL} \url{http://dx.doi.org/10.1007/s13347-023-00687-8} \end{APACrefURL}
\newblock
\begin{APACrefDOI} \doi{10.1007/s13347-023-00687-8} \end{APACrefDOI}
\PrintBackRefs{\CurrentBib}

\bibitem [\protect \citeauthoryear {%
Munoz%
, Sawyer%
\BCBL {}\ \BBA {} Dunn%
}{%
Munoz%
\ \protect \BOthers {.}}{%
{\protect \APACyear {2022}}%
}]{%
Munoz2022}
\APACinsertmetastar {%
Munoz2022}%
\begin{APACrefauthors}%
Munoz, I.%
, Sawyer, S.%
\BCBL {}\ \BBA {} Dunn, M.%
\end{APACrefauthors}%
\unskip\
\newblock
\APACrefYearMonthDay{2022}{}{}.
\newblock
{\BBOQ}\APACrefatitle {New futures of work or continued marginalization? The rise of online freelance work and digital platforms} {New futures of work or continued marginalization? the rise of online freelance work and digital platforms}.{\BBCQ}
\newblock
\BIn{} \APACrefbtitle {Proceedings of the 1st Annual Meeting of the Symposium on Human-Computer Interaction for Work.} {Proceedings of the 1st annual meeting of the symposium on human-computer interaction for work.}
\newblock
\APACaddressPublisher{New York, NY, USA}{Association for Computing Machinery}.
\newblock
\begin{APACrefURL} \url{https://doi.org/10.1145/3533406.3533412} \end{APACrefURL}
\newblock
\begin{APACrefDOI} \doi{10.1145/3533406.3533412} \end{APACrefDOI}
\PrintBackRefs{\CurrentBib}

\bibitem [\protect \citeauthoryear {%
Negri%
}{%
Negri%
}{%
{\protect \APACyear {1979}}%
}]{%
Negri1979}
\APACinsertmetastar {%
Negri1979}%
\begin{APACrefauthors}%
Negri, A.%
\end{APACrefauthors}%
\unskip\
\newblock
\APACrefYear{1979}.
\newblock
\APACrefbtitle {Marx oltre Marx: quaderno di lavoro sui Grundrisse} {Marx oltre marx: quaderno di lavoro sui grundrisse}.
\newblock
\APACaddressPublisher{}{Feltrinelli}.
\PrintBackRefs{\CurrentBib}

\bibitem [\protect \citeauthoryear {%
Papadimitropoulos%
}{%
Papadimitropoulos%
}{%
{\protect \APACyear {2021}}%
}]{%
Papadimitropoulos2021}
\APACinsertmetastar {%
Papadimitropoulos2021}%
\begin{APACrefauthors}%
Papadimitropoulos, E.%
\end{APACrefauthors}%
\unskip\
\newblock
\APACrefYearMonthDay{2021}{{\APACmonth{04}}}{}.
\newblock
{\BBOQ}\APACrefatitle {Platform capitalism, platform cooperativism, and the commons} {Platform capitalism, platform cooperativism, and the commons}.{\BBCQ}
\newblock
\APACjournalVolNumPages{Rethink. Marx.}{33}{2}{246--262}.
\PrintBackRefs{\CurrentBib}

\bibitem [\protect \citeauthoryear {%
Pianta%
}{%
Pianta%
}{%
{\protect \APACyear {2020}}%
}]{%
Pianta2020}
\APACinsertmetastar {%
Pianta2020}%
\begin{APACrefauthors}%
Pianta, M.%
\end{APACrefauthors}%
\unskip\
\newblock
\APACrefYearMonthDay{2020}{}{}.
\newblock
{\BBOQ}\APACrefatitle {Technology and Work: Key Stylized Facts for the Digital Age} {Technology and work: Key stylized facts for the digital age}.{\BBCQ}
\newblock
\BIn{} \APACrefbtitle {Handbook of Labor, Human Resources and Population Economics} {Handbook of labor, human resources and population economics}\ (\BPG~1–17).
\newblock
\APACaddressPublisher{}{Springer International Publishing}.
\newblock
\begin{APACrefDOI} \doi{10.1007/978-3-319-57365-6_3-1} \end{APACrefDOI}
\PrintBackRefs{\CurrentBib}

\bibitem [\protect \citeauthoryear {%
Rani%
, Castel-Branco%
, Satija%
\BCBL {}\ \BBA {} Nayar%
}{%
Rani%
\ \protect \BOthers {.}}{%
{\protect \APACyear {2022}}%
}]{%
Rani2022}
\APACinsertmetastar {%
Rani2022}%
\begin{APACrefauthors}%
Rani, U.%
, Castel-Branco, R.%
, Satija, S.%
\BCBL {}\ \BBA {} Nayar, M.%
\end{APACrefauthors}%
\unskip\
\newblock
\APACrefYearMonthDay{2022}{}{}.
\newblock
{\BBOQ}\APACrefatitle {Women, work, and the digital economy} {Women, work, and the digital economy}.{\BBCQ}
\newblock
\APACjournalVolNumPages{Gender \& Development}{30}{3}{421--435}.
\newblock
\begin{APACrefDOI} \doi{10.1080/13552074.2022.2151729} \end{APACrefDOI}
\PrintBackRefs{\CurrentBib}

\bibitem [\protect \citeauthoryear {%
Ricaurte%
}{%
Ricaurte%
}{%
{\protect \APACyear {2019}}%
}]{%
Ricaurte2019}
\APACinsertmetastar {%
Ricaurte2019}%
\begin{APACrefauthors}%
Ricaurte, P.%
\end{APACrefauthors}%
\unskip\
\newblock
\APACrefYearMonthDay{2019}{}{}.
\newblock
{\BBOQ}\APACrefatitle {Data Epistemologies, The Coloniality of Power, and Resistance} {Data epistemologies, the coloniality of power, and resistance}.{\BBCQ}
\newblock
\APACjournalVolNumPages{Television \& New Media}{20}{4}{350-365}.
\newblock
\begin{APACrefDOI} \doi{10.1177/1527476419831640} \end{APACrefDOI}
\PrintBackRefs{\CurrentBib}

\bibitem [\protect \citeauthoryear {%
Richardson%
}{%
Richardson%
}{%
{\protect \APACyear {2016}}%
}]{%
Richardson2016}
\APACinsertmetastar {%
Richardson2016}%
\begin{APACrefauthors}%
Richardson, L.%
\end{APACrefauthors}%
\unskip\
\newblock
\APACrefYearMonthDay{2016}{{\APACmonth{11}}}{}.
\newblock
{\BBOQ}\APACrefatitle {Feminist geographies of digital work} {Feminist geographies of digital work}.{\BBCQ}
\newblock
\APACjournalVolNumPages{Progress in Human Geography}{42}{2}{244–263}.
\newblock
\begin{APACrefDOI} \doi{10.1177/0309132516677177} \end{APACrefDOI}
\PrintBackRefs{\CurrentBib}

\bibitem [\protect \citeauthoryear {%
Rifkin%
}{%
Rifkin%
}{%
{\protect \APACyear {1995}}%
}]{%
Rifkin1995}
\APACinsertmetastar {%
Rifkin1995}%
\begin{APACrefauthors}%
Rifkin, J.%
\end{APACrefauthors}%
\unskip\
\newblock
\APACrefYear{1995}.
\newblock
\APACrefbtitle {The end of work. The Decline ofthe Global Labor Force and the Dawn ofthe Post-Market Era} {The end of work. the decline ofthe global labor force and the dawn ofthe post-market era}.
\newblock
\APACaddressPublisher{New York, NY}{Tarcher/Putnam}.
\PrintBackRefs{\CurrentBib}

\bibitem [\protect \citeauthoryear {%
Rizzo%
}{%
Rizzo%
}{%
{\protect \APACyear {2023}}%
}]{%
Rizzo2023}
\APACinsertmetastar {%
Rizzo2023}%
\begin{APACrefauthors}%
Rizzo, J.%
\end{APACrefauthors}%
\unskip\
\newblock
\APACrefYearMonthDay{2023}{}{}.
\newblock
{\BBOQ}\APACrefatitle {Asking for It: Gendered Dimensions of Surveillance Capitalism} {Asking for it: Gendered dimensions of surveillance capitalism}.{\BBCQ}
\newblock
\APACjournalVolNumPages{Emancipations}{}{}{}.
\PrintBackRefs{\CurrentBib}

\bibitem [\protect \citeauthoryear {%
Rodríguez-Modroño%
, Pesole%
\BCBL {}\ \BBA {} López-Igual%
}{%
Rodríguez-Modroño%
\ \protect \BOthers {.}}{%
{\protect \APACyear {2022}}%
}]{%
Rodriguez2022}
\APACinsertmetastar {%
Rodriguez2022}%
\begin{APACrefauthors}%
Rodríguez-Modroño, P.%
, Pesole, A.%
\BCBL {}\ \BBA {} López-Igual, P.%
\end{APACrefauthors}%
\unskip\
\newblock
\APACrefYearMonthDay{2022}{}{}.
\newblock
{\BBOQ}\APACrefatitle {Assessing gender inequality in digital labour platforms in Europe} {Assessing gender inequality in digital labour platforms in europe}.{\BBCQ}
\newblock
\APACjournalVolNumPages{Internet Policy Review}{11}{1}{1--23}.
\newblock
\APACrefnote{Publisher: Berlin: Alexander von Humboldt Institute for Internet and Society}
\newblock
\begin{APACrefDOI} \doi{10.14763/2022.1.1622} \end{APACrefDOI}
\PrintBackRefs{\CurrentBib}

\bibitem [\protect \citeauthoryear {%
Scholz%
}{%
Scholz%
}{%
{\protect \APACyear {2016}}%
}]{%
Scholz2016b}
\APACinsertmetastar {%
Scholz2016b}%
\begin{APACrefauthors}%
Scholz, T.%
\end{APACrefauthors}%
\unskip\
\newblock
\APACrefYearMonthDay{2016}{}{}.
\newblock
\APACrefbtitle {Platform Cooperativism. Challenging the Corporate Sharing Economy} {Platform cooperativism. challenging the corporate sharing economy}\ \APACbVolEdTR{}{\BTR{}}.
\newblock
\APACaddressInstitution{}{New York: Rosa Luxemburg Stiftung}.
\PrintBackRefs{\CurrentBib}

\bibitem [\protect \citeauthoryear {%
Scholz%
\ \BBA {} Schneider%
}{%
Scholz%
\ \BBA {} Schneider%
}{%
{\protect \APACyear {2016}}%
}]{%
Scholz2016a}
\APACinsertmetastar {%
Scholz2016a}%
\begin{APACrefauthors}%
Scholz, T.%
\BCBT {}\ \BBA {} Schneider, N.%
\end{APACrefauthors}%
\ (\BEDS).
\unskip\
\newblock
\APACrefYear{2016}.
\newblock
\APACrefbtitle {Ours to hack and to own: The Rise of Platform Cooperativism, A New Vision for the Future of Work and a Fairer Internet} {Ours to hack and to own: The rise of platform cooperativism, a new vision for the future of work and a fairer internet}.
\newblock
\APACaddressPublisher{}{OR Books}.
\PrintBackRefs{\CurrentBib}

\bibitem [\protect \citeauthoryear {%
Schumpeter%
}{%
Schumpeter%
}{%
{\protect \APACyear {1976}}%
}]{%
Schumpeter1976}
\APACinsertmetastar {%
Schumpeter1976}%
\begin{APACrefauthors}%
Schumpeter, J.%
\end{APACrefauthors}%
\unskip\
\newblock
\APACrefYear{1976}.
\newblock
\APACrefbtitle {Capitalism, Socialism and Democracy} {Capitalism, socialism and democracy}.
\newblock
\APACaddressPublisher{}{Routledge}.
\newblock
\begin{APACrefURL} \url{https://books.google.it/books?id=6eM6YrMj46sC} \end{APACrefURL}
\PrintBackRefs{\CurrentBib}

\bibitem [\protect \citeauthoryear {%
Sch\"{u}rmann%
\ \BBA {} Trenkmann%
}{%
Sch\"{u}rmann%
\ \BBA {} Trenkmann%
}{%
{\protect \APACyear {2023}}%
}]{%
Schurmann2023}
\APACinsertmetastar {%
Schurmann2023}%
\begin{APACrefauthors}%
Sch\"{u}rmann, L.%
\BCBT {}\ \BBA {} Trenkmann, J.%
\end{APACrefauthors}%
\unskip\
\newblock
\APACrefYearMonthDay{2023}{}{}.
\newblock
{\BBOQ}\APACrefatitle {Rethinking digital work arrangements from a feminist perspective. The ``cyborg'' as the epitome of hybrid working} {Rethinking digital work arrangements from a feminist perspective. the ``cyborg'' as the epitome of hybrid working}.{\BBCQ}
\newblock
\APACjournalVolNumPages{Rassegna Italiana di Sociologia}{}{}{261–286}.
\newblock
\begin{APACrefDOI} \doi{10.1423/107860} \end{APACrefDOI}
\PrintBackRefs{\CurrentBib}

\bibitem [\protect \citeauthoryear {%
A.~Smith%
}{%
A.~Smith%
}{%
{\protect \APACyear {2015}}%
}]{%
Smith2015}
\APACinsertmetastar {%
Smith2015}%
\begin{APACrefauthors}%
Smith, A.%
\end{APACrefauthors}%
\unskip\
\newblock
\APACrefYearMonthDay{2015}{}{}.
\newblock
{\BBOQ}\APACrefatitle {Not-Seeing: State Surveillance, Settler Colonialism, and Gender Violence} {Not-seeing: State surveillance, settler colonialism, and gender violence}.{\BBCQ}
\newblock
\BIn{} \APACrefbtitle {Feminist Surveillance Studies} {Feminist surveillance studies}\ (\BPG~21–38).
\newblock
\APACaddressPublisher{}{Duke University Press}.
\newblock
\begin{APACrefDOI} \doi{10.1215/9780822375463-002} \end{APACrefDOI}
\PrintBackRefs{\CurrentBib}

\bibitem [\protect \citeauthoryear {%
G.~Smith%
\ \BBA {} Rustagi%
}{%
G.~Smith%
\ \BBA {} Rustagi%
}{%
{\protect \APACyear {2021}}%
}]{%
smith2021good}
\APACinsertmetastar {%
smith2021good}%
\begin{APACrefauthors}%
Smith, G.%
\BCBT {}\ \BBA {} Rustagi, I.%
\end{APACrefauthors}%
\unskip\
\newblock
\APACrefYearMonthDay{2021}{}{}.
\newblock
{\BBOQ}\APACrefatitle {When good algorithms go sexist: Why and how to advance AI gender equity} {When good algorithms go sexist: Why and how to advance ai gender equity}.{\BBCQ}
\newblock
\APACjournalVolNumPages{Standford Social Innovation Review}{}{}{1--8}.
\PrintBackRefs{\CurrentBib}

\bibitem [\protect \citeauthoryear {%
Stefano%
}{%
Stefano%
}{%
{\protect \APACyear {2015}}%
}]{%
Stefano2015}
\APACinsertmetastar {%
Stefano2015}%
\begin{APACrefauthors}%
Stefano, V\BPBI D.%
\end{APACrefauthors}%
\unskip\
\newblock
\APACrefYearMonthDay{2015}{}{}.
\newblock
{\BBOQ}\APACrefatitle {The rise of the "just-in-time workforce": on-demand work, crowdwork and labour protection in the "gig-economy"} {The rise of the "just-in-time workforce": on-demand work, crowdwork and labour protection in the "gig-economy"}.{\BBCQ}
\newblock
\APACjournalVolNumPages{Comparative Labor Law and Policy Journal}{37}{}{461-471}.
\newblock
\begin{APACrefURL} \url{https://api.semanticscholar.org/CorpusID:154451333} \end{APACrefURL}
\PrintBackRefs{\CurrentBib}

\bibitem [\protect \citeauthoryear {%
Surie%
\ \BBA {} Huws%
}{%
Surie%
\ \BBA {} Huws%
}{%
{\protect \APACyear {2023}}%
}]{%
Surie2023}
\APACinsertmetastar {%
Surie2023}%
\begin{APACrefauthors}%
Surie, A.%
\BCBT {}\ \BBA {} Huws, U.%
\end{APACrefauthors}%
\unskip\
\newblock
\APACrefYearMonthDay{2023}{}{}.
\newblock
{\BBOQ}\APACrefatitle {Platformization and Informality: Pathways of Change, Alteration, and Transformation} {Platformization and informality: Pathways of change, alteration, and transformation}.{\BBCQ}
\newblock
\BIn{} \APACrefbtitle {Platformization and Informality} {Platformization and informality}\ (\BPG~1–12).
\newblock
\APACaddressPublisher{}{Springer International Publishing}.
\newblock
\begin{APACrefDOI} \doi{10.1007/978-3-031-11462-5_1} \end{APACrefDOI}
\PrintBackRefs{\CurrentBib}

\bibitem [\protect \citeauthoryear {%
Terranova%
}{%
Terranova%
}{%
{\protect \APACyear {2012}}%
}]{%
Terranova2012}
\APACinsertmetastar {%
Terranova2012}%
\begin{APACrefauthors}%
Terranova, T.%
\end{APACrefauthors}%
\unskip\
\newblock
\APACrefYearMonthDay{2012}{}{}.
\newblock
{\BBOQ}\APACrefatitle {Free labor} {Free labor}.{\BBCQ}
\newblock
\BIn{} \APACrefbtitle {Digital Labor} {Digital labor}\ (\BPGS\ 33--57).
\newblock
\APACaddressPublisher{}{Routledge}.
\PrintBackRefs{\CurrentBib}

\bibitem [\protect \citeauthoryear {%
{The European Parliament and the Council of the European Union}%
}{%
{The European Parliament and the Council of the European Union}%
}{%
{\protect \APACyear {2024}}%
}]{%
EUDirectiveWork2024}
\APACinsertmetastar {%
EUDirectiveWork2024}%
\begin{APACrefauthors}%
{The European Parliament and the Council of the European Union}.%
\end{APACrefauthors}%
\unskip\
\newblock
\APACrefYearMonthDay{2024}{}{}.
\newblock
\APACrefbtitle {Proposal for the ''Directive of the European Parliamente and the Council'' on improving working conditions in platform work.} {Proposal for the ''directive of the european parliamente and the council'' on improving working conditions in platform work.}
\newblock
\APACrefnote{Document 2021/0414(COD)}
\PrintBackRefs{\CurrentBib}

\bibitem [\protect \citeauthoryear {%
Tong%
}{%
Tong%
}{%
{\protect \APACyear {2018}}%
}]{%
Tong2018}
\APACinsertmetastar {%
Tong2018}%
\begin{APACrefauthors}%
Tong, R.%
\end{APACrefauthors}%
\unskip\
\newblock
\APACrefYear{2018}.
\newblock
\APACrefbtitle {Feminist thought: A more comprehensive introduction} {Feminist thought: A more comprehensive introduction}.
\newblock
\APACaddressPublisher{}{Routledge}.
\PrintBackRefs{\CurrentBib}

\bibitem [\protect \citeauthoryear {%
Tronti%
}{%
Tronti%
}{%
{\protect \APACyear {2019}}%
}]{%
Tronti2019}
\APACinsertmetastar {%
Tronti2019}%
\begin{APACrefauthors}%
Tronti, M.%
\end{APACrefauthors}%
\unskip\
\newblock
\APACrefYear{2019}.
\newblock
\APACrefbtitle {Workers and capital} {Workers and capital}.
\newblock
\APACaddressPublisher{}{Verso Books}.
\PrintBackRefs{\CurrentBib}

\bibitem [\protect \citeauthoryear {%
Trotsky%
}{%
Trotsky%
}{%
{\protect \APACyear {1977}}%
}]{%
Trotsky1977}
\APACinsertmetastar {%
Trotsky1977}%
\begin{APACrefauthors}%
Trotsky, L.%
\end{APACrefauthors}%
\unskip\
\newblock
\APACrefYear{1977}.
\newblock
\APACrefbtitle {The History of the Russian revolution} {The history of the russian revolution}.
\newblock
\APACaddressPublisher{London, England}{Pluto Press}.
\newblock
\APACrefnote{Translated by Max Eastman}
\PrintBackRefs{\CurrentBib}

\bibitem [\protect \citeauthoryear {%
Tubaro%
, Coville%
, Le~Ludec%
\BCBL {}\ \BBA {} Casilli%
}{%
Tubaro%
\ \protect \BOthers {.}}{%
{\protect \APACyear {2022}}%
}]{%
Tubaro2022}
\APACinsertmetastar {%
Tubaro2022}%
\begin{APACrefauthors}%
Tubaro, P.%
, Coville, M.%
, Le~Ludec, C.%
\BCBL {}\ \BBA {} Casilli, A\BPBI A.%
\end{APACrefauthors}%
\unskip\
\newblock
\APACrefYearMonthDay{2022}{{\APACmonth{02}}}{}.
\newblock
{\BBOQ}\APACrefatitle {Hidden inequalities: the gendered labour of women on micro-tasking platforms} {Hidden inequalities: the gendered labour of women on micro-tasking platforms}.{\BBCQ}
\newblock
\APACjournalVolNumPages{Internet Policy Review}{11}{1}{}.
\newblock
\begin{APACrefURL} \url{http://dx.doi.org/10.14763/2022.1.1623} \end{APACrefURL}
\newblock
\begin{APACrefDOI} \doi{10.14763/2022.1.1623} \end{APACrefDOI}
\PrintBackRefs{\CurrentBib}

\bibitem [\protect \citeauthoryear {%
{UN Women}%
}{%
{UN Women}%
}{%
{\protect \APACyear {2020}}%
}]{%
UNWomen2020}
\APACinsertmetastar {%
UNWomen2020}%
\begin{APACrefauthors}%
{UN Women}.%
\end{APACrefauthors}%
\unskip\
\newblock
\APACrefYearMonthDay{2020}{}{}.
\newblock
\APACrefbtitle {The digital revolution: Implications for gender equality and women’s rights 25 years after Beijing.} {The digital revolution: Implications for gender equality and women’s rights 25 years after beijing.}
\PrintBackRefs{\CurrentBib}

\bibitem [\protect \citeauthoryear {%
van Doorn%
\ \BBA {} Vijay%
}{%
van Doorn%
\ \BBA {} Vijay%
}{%
{\protect \APACyear {2024}}%
}]{%
vanDoorn2024}
\APACinsertmetastar {%
vanDoorn2024}%
\begin{APACrefauthors}%
van Doorn, N.%
\BCBT {}\ \BBA {} Vijay, D.%
\end{APACrefauthors}%
\unskip\
\newblock
\APACrefYearMonthDay{2024}{}{}.
\newblock
{\BBOQ}\APACrefatitle {Gig work as migrant work: The platformization of migration infrastructure} {Gig work as migrant work: The platformization of migration infrastructure}.{\BBCQ}
\newblock
\APACjournalVolNumPages{Environment and Planning A: Economy and Space}{56}{4}{1129-1149}.
\newblock
\begin{APACrefDOI} \doi{10.1177/0308518X211065049} \end{APACrefDOI}
\PrintBackRefs{\CurrentBib}

\bibitem [\protect \citeauthoryear {%
Wang%
\ \BBA {} Tomassetti%
}{%
Wang%
\ \BBA {} Tomassetti%
}{%
{\protect \APACyear {2024}}%
}]{%
Wang2024}
\APACinsertmetastar {%
Wang2024}%
\begin{APACrefauthors}%
Wang, J.%
\BCBT {}\ \BBA {} Tomassetti, J.%
\end{APACrefauthors}%
\unskip\
\newblock
\APACrefYearMonthDay{2024}{{\APACmonth{02}}}{}.
\newblock
{\BBOQ}\APACrefatitle {Labor-capital relations on digital platforms: Organization, algorithmic discipline and the social factory again} {Labor-capital relations on digital platforms: Organization, algorithmic discipline and the social factory again}.{\BBCQ}
\newblock
\APACjournalVolNumPages{Sociology Compass}{18}{3}{}.
\newblock
\begin{APACrefDOI} \doi{10.1111/soc4.13192} \end{APACrefDOI}
\PrintBackRefs{\CurrentBib}

\bibitem [\protect \citeauthoryear {%
Webster%
\ \BBA {} Zhang%
}{%
Webster%
\ \BBA {} Zhang%
}{%
{\protect \APACyear {2020}}%
}]{%
Webster2020}
\APACinsertmetastar {%
Webster2020}%
\begin{APACrefauthors}%
Webster, N\BPBI A.%
\BCBT {}\ \BBA {} Zhang, Q.%
\end{APACrefauthors}%
\unskip\
\newblock
\APACrefYearMonthDay{2020}{{\APACmonth{02}}}{}.
\newblock
{\BBOQ}\APACrefatitle {Careers Delivered from the Kitchen? Immigrant Women Small-scale Entrepreneurs Working in the Growing Nordic Platform Economy} {Careers delivered from the kitchen? immigrant women small-scale entrepreneurs working in the growing nordic platform economy}.{\BBCQ}
\newblock
\APACjournalVolNumPages{NORA - Nordic Journal of Feminist and Gender Research}{28}{2}{113–125}.
\newblock
\begin{APACrefDOI} \doi{10.1080/08038740.2020.1714725} \end{APACrefDOI}
\PrintBackRefs{\CurrentBib}

\bibitem [\protect \citeauthoryear {%
Webster%
\ \BBA {} Zhang%
}{%
Webster%
\ \BBA {} Zhang%
}{%
{\protect \APACyear {2022}}%
}]{%
Webster2022}
\APACinsertmetastar {%
Webster2022}%
\begin{APACrefauthors}%
Webster, N\BPBI A.%
\BCBT {}\ \BBA {} Zhang, Q.%
\end{APACrefauthors}%
\unskip\
\newblock
\APACrefYearMonthDay{2022}{}{}.
\newblock
{\BBOQ}\APACrefatitle {Intersectional understandings of the role and meaning of platform-mediated work in the pandemic Swedish welfare state} {Intersectional understandings of the role and meaning of platform-mediated work in the pandemic swedish welfare state}.{\BBCQ}
\newblock
\APACjournalVolNumPages{Digital Geography and Society}{3}{}{100025}.
\newblock
\begin{APACrefDOI} \doi{10.1016/j.diggeo.2021.100025} \end{APACrefDOI}
\PrintBackRefs{\CurrentBib}

\bibitem [\protect \citeauthoryear {%
Weeks%
}{%
Weeks%
}{%
{\protect \APACyear {2011}}%
}]{%
Weeks2011}
\APACinsertmetastar {%
Weeks2011}%
\begin{APACrefauthors}%
Weeks, K.%
\end{APACrefauthors}%
\unskip\
\newblock
\APACrefYear{2011}.
\newblock
\APACrefbtitle {The Problem with Work: Feminism, Marxism, Antiwork Politics, and Postwork Imaginaries} {The problem with work: Feminism, marxism, antiwork politics, and postwork imaginaries}.
\newblock
\APACaddressPublisher{}{Duke University Press}.
\PrintBackRefs{\CurrentBib}

\bibitem [\protect \citeauthoryear {%
Zuboff%
}{%
Zuboff%
}{%
{\protect \APACyear {2014}}%
}]{%
Zuboff2014}
\APACinsertmetastar {%
Zuboff2014}%
\begin{APACrefauthors}%
Zuboff, S.%
\end{APACrefauthors}%
\unskip\
\newblock
\APACrefYearMonthDay{2014}{}{}.
\newblock
{\BBOQ}\APACrefatitle {A digital declaration: Big data as surveillance capitalism} {A digital declaration: Big data as surveillance capitalism}.{\BBCQ}
\newblock
\APACjournalVolNumPages{FAZ .NET}{}{}{}.
\PrintBackRefs{\CurrentBib}

\bibitem [\protect \citeauthoryear {%
Zuboff%
}{%
Zuboff%
}{%
{\protect \APACyear {2019}}%
{\protect \APACexlab {{\protect \BCnt {1}}}}}]{%
Zuboff2019}
\APACinsertmetastar {%
Zuboff2019}%
\begin{APACrefauthors}%
Zuboff, S.%
\end{APACrefauthors}%
\unskip\
\newblock
\APACrefYear{2019{\protect \BCnt {1}}}.
\newblock
\APACrefbtitle {The age of surveillance capitalism} {The age of surveillance capitalism}.
\newblock
\APACaddressPublisher{London, England}{Profile Books}.
\PrintBackRefs{\CurrentBib}

\bibitem [\protect \citeauthoryear {%
Zuboff%
}{%
Zuboff%
}{%
{\protect \APACyear {2019}}%
{\protect \APACexlab {{\protect \BCnt {2}}}}}]{%
Zuboff2019b}
\APACinsertmetastar {%
Zuboff2019b}%
\begin{APACrefauthors}%
Zuboff, S.%
\end{APACrefauthors}%
\unskip\
\newblock
\APACrefYearMonthDay{2019{\protect \BCnt {2}}}{}{}.
\newblock
{\BBOQ}\APACrefatitle {Surveillance Capitalism and the Challenge of Collective Action} {Surveillance capitalism and the challenge of collective action}.{\BBCQ}
\newblock
\APACjournalVolNumPages{New Labor Forum}{28}{1}{10-29}.
\newblock
\begin{APACrefDOI} \doi{10.1177/1095796018819461} \end{APACrefDOI}
\PrintBackRefs{\CurrentBib}

\bibitem [\protect \citeauthoryear {%
Zuboff%
}{%
Zuboff%
}{%
{\protect \APACyear {2020}}%
}]{%
Zuboff2020}
\APACinsertmetastar {%
Zuboff2020}%
\begin{APACrefauthors}%
Zuboff, S.%
\end{APACrefauthors}%
\unskip\
\newblock
\APACrefYearMonthDay{2020}{}{}.
\newblock
\APACrefbtitle {You Are Now Remotely Controlled.} {You are now remotely controlled.}
\newblock
\begin{APACrefURL} \url{https://www.nytimes.com/2020/01/24/opinion/sunday/surveillance-capitalism.html} \end{APACrefURL}
\PrintBackRefs{\CurrentBib}

\end{thebibliography}

\end{document}